\documentclass[%
iicol,
pdflatex,
]{sn-jnl}

\usepackage{silence}
\usepackage{graphicx}
\usepackage{booktabs}
\usepackage{dcolumn}
\usepackage{amsmath,amssymb,amsfonts}%
\usepackage{bm}
\usepackage{makecell}
\usepackage{float}
\usepackage[T1]{fontenc}
\usepackage{xcolor}
\usepackage{textcomp,gensymb} 
\usepackage{soul}
\usepackage{rotating}
\usepackage{chngcntr}
\counterwithin*{figure}{section}
\usepackage[super,sort&compress]{natbib}
\usepackage{aas_macros}
\usepackage{caption}
\renewcommand{\figurename}{{\bf Figure}}
\renewcommand{\thefigure}{{\bf \arabic{figure}}}

\renewcommand{\tablename}{{\bf Table}}
\renewcommand{\thetable}{{\bf \arabic{table}}}

\usepackage{hyperref} 
\hypersetup{
    colorlinks=true,
    linkcolor=blue,
    filecolor=magenta,      
    urlcolor=cyan,
}
\begin{document}

\title{Efficient ion re-acceleration in laboratory-produced interpenetrating collisionless shocks}

\author[1,2]{W. Yao}
\author[1,3]{I. Cohen}
\author[1]{P. Suarez Gerona}
\author[4]{H. Ahmed}
\author[5]{A.F.A. Bott}
\author[6]{S. N. Chen}
\author[4]{M. Cook}
\author[1,7]{R. Leli\`evre}
\author[8]{P. Martin}
\author[1,9,10]{T. Waltenspiel}
\author[10]{P. Antici}
\author[11]{J. B\'eard}
\author[8]{M. Borghesi}
\author[12,13]{D. Caprioli}
\author[2]{A. Ciardi}
\author[9]{E. d'Humi\`eres}
\author[9]{M. François}
\author[14,15]{L. Gremillet}
\author[16]{A. Marcowith}
\author[17]{M. Miceli}
\author[1,2,18]{T. Seebaruth}
\author[19]{S. Orlando}
\author[1,3]{J. Fuchs}

\affil[1]{LULI - CNRS, CEA, UPMC Univ Paris 06 : Sorbonne Universit\'e, Ecole Polytechnique, Institut Polytechnique de Paris - F-91128 Palaiseau cedex, France}
\affil[2]{Sorbonne Université, Observatoire de Paris, Université PSL, Laboratoire d'étude de l'Univers et des phénomènes eXtrêmes, LUX, CNRS, F-75005 Meudon, France}
\affil[3]{\orgdiv{Department of Physics}, \orgname{Technion}, \orgaddress{\city{Haifa}, \postcode{32000}, \country{Israel}}}
\affil[4]{Central Laser Facility, STFC Rutherford Appleton Laboratory, Oxfordshire OX11 0QX, United Kingdom}
\affil[5]{Department of Physics, University of Oxford, Parks Road, Oxford OX1 3PU, United Kingdom}
\affil[6]{ELI-NP, "Horia Hulubei" National Institute for Physics and Nuclear Engineering, 30 Reactorului Street, RO-077125, Bucharest-Magurele, Romania}
\affil[7]{Laboratoire de micro-irradiation, de métrologie et de dosimétrie des neutrons, PSE-Santé/SDOS, IRSN, 13115 Saint-Paul-Lez-Durance, France}
\affil[8]{Center for Plasma Physics, School of Mathematics and Physics, Queen's University Belfast, Belfast BT7 1NN, United Kingdom}
\affil[9]{University of Bordeaux, Centre Lasers Intenses et Applications, CNRS, CEA, UMR 5107, F-33405 Talence, France}
\affil[10]{INRS-EMT, 1650 boul, Lionel-Boulet, Varennes, QC, J3X 1S2, Canada}
\affil[11]{LNCMI, UPR 3228, CNRS-UGA-UPS-INSA, Toulouse 31400, France}
\affil[12]{Department of Astronomy \& Astrophysics, University of Chicago, Chicago, Illinois 60637, USA}
\affil[13]{Enrico Fermi Institute, The University of Chicago, Chicago, Illinois 60637, USA}
\affil[14]{CEA, DAM, DIF, F-91297 Arpajon, France}
\affil[15]{Université Paris-Saclay, CEA, LMCE, F-91680 Bruyères-le-Châtel, France}
\affil[16]{Laboratoire Univers et Particules de Montpellier CNRS/Université de Montpellier, Place E. Bataillon, 34095 Montpellier, France}
\affil[17]{University of Palermo, Department of Physics and Chemistry, Palermo, Italy}
\affil[18]{Laboratoire de Physique des Plasmas (LPP), CNRS, Observatoire de Paris, Sorbonne Université, Université Paris-Saclay, École polytechnique, Institut Polytechnique de
Paris, F-91120 Palaiseau, France}
\affil[19]{INAF–Osservatorio Astronomico di Palermo, Palermo, Italy}

\date{\today}

\maketitle


\textbf{Although the origin of cosmic rays (CRs) remains an open question, collisionless magnetized shock waves are widely regarded as key sites for particle acceleration \cite{drury2012origin}. Recent theories further suggest that shock-shock collisions  in stellar clusters \cite{2004A&A...424..747P, 2001SSRv...99..317B} could provide the additional acceleration needed to explain the observed high-energy CR spectrum \cite{bykov2013non
,2020MNRAS.494.3166V, 2023PhRvL.131i5201O, 2023PhRvE.107b5201M}. Here, we investigate this hypothesis through a controlled laser-based experiment, where we create magnetized plasma conditions similar to astrophysical environments. Our results demonstrate that interpenetrating collisionless shocks can significantly boost the energy of ambient protons previously energized by the individual shocks by the field structure, thus improving the overall acceleration efficiency. 
Numerical kinetic simulations corroborate these findings, revealing that protons are reaccelerated via their bouncing motion in the convective electric fields of the colliding magnetized flows. \textcolor{red}{More generally, }our novel experimental platform \textcolor{red}{enables quantitative evaluation of supercritical magnetized shocks interactions with the ambient plasma and of the associated particle energization processes, } opening the prospect to test other \textcolor{red}{theoretical} mechanisms in a controlled laboratory setting. }

Single collisionless shocks have been extensively studied regarding their prospects as particle accelerators, however, the existence of multiple interacting shocks from supernova (SN) events is far greater.  One interesting locale where shock-shock collisions can be found is indeed in clusters of hot giant stars, where frequent and correlated SN explosions take place \cite{higdon1998cosmic}.  The interplay between stellar winds and SN events in these regions gives rise to large-scale structures known as ``superbubbles'' \cite{ackermann2011cocoon}, within which various  supercritical\footnote{I.e., shocks with a magnetosonic Mach number $M_{ms} = v_{sh}/\sqrt{v_A^2+c_s^2}$ exceeding, in a MHD framework and for perpendicular shocks, the critical value of 2.7\cite{balogh2013physics} over which ion are reflected by the shock - here, $v_{sh}$, $v_A$ and $c_s$ are the shock, ambient Alfvénic and sound velocities, respectively.} shock-shock collisions occur, including wind-wind, SN-wind, and SN-SN shock interactions \cite{bykov2013non,bykov2015ultrahard}. 

Several theoretical models have attempted to understand the acceleration of cosmic rays (CRs) in colliding-shock configurations. For instance, Vieu \emph{et al.} \cite{2020MNRAS.494.3166V} proposed that, although the fundamental acceleration process remains unchanged when two shocks collide, spectral hardening can occur for the highest-energy particles that are scattered between the two shock fronts. Complementarily, Malkov and Lemoine  \cite{2023PhRvE.107b5201M} suggested that converging perpendicular shocks could efficiently re-accelerate preexisting CRs. 
\textcolor{red}{Though we will not directly test such models, since our setup cannot reproduce the modeled configurations; nonetheless, we will show that our laboratory setup is able to}
\textcolor{red}{establish a significant }
re-acceleration of ions 
\textcolor{red}{is induced by} 
colliding supercritical shocks.

\textcolor{red}{
Here, we focus on supercritical perpendicular shocks -- where the background magnetic field is perpendicular to the shock normal -- which can be produced in laboratory setups \cite{schaeffer2019direct,yao2021laboratory}. Given that astrophysical shocks generally propagate through turbulent media \cite{Trotta2023} and are often corrugated, the magnetic field obliquity at the shock front is inherently random. Our configuration thus aims at mimicking part of the 3D front of supercritical shocks found in astrophysical environments.} \textcolor{red}{Note that our setup also does not specifically involve turbulence, as we try here to specifically highlight the ion energy boost by electric fields of the interpenetrating shocks.} 

Our experimental setup (see Methods) investigates ion acceleration from two converging, perpendicular, supercritical shocks. These shocks, as their astrophysical counterparts, are ``collisionless'' because the width of their transition layer, given by the ion inertial length $d_i$ \cite{balogh2013physics}, is much smaller than the collisional mean-free-path ($\lambda_{mfp}$) of the ions (see Table~\ref{tab:parameters}, \textcolor{red}{and Section 4 of the Supp. Info.}). Instead of Coulomb collisions, shock formation is mediated by wave-particle interactions in self-induced or pre-existing electromagnetic fields. Also note that in our conditions, as in astrophysical systems, the ion Larmor radius of upstream protons encountering the shock  is much larger than the shock width. Such structures are ubiquitous phenomena and have been observed across a wide range of scales \cite{balogh2013physics}, from astrophysical ($10^{16}\sim 10^{25}$ m) \cite{helder2009measuring}, to space ($10^{8}\sim 10^{12}$ m) \cite{turner2018autogenous} plasmas, and down to laboratory environments ($10^{-6}\sim 10^{-3}$ m) \cite{fiuza2020electron,yao2021laboratory}.

\begin{figure*}[htp]
    \centering
    \includegraphics[width=1.0\textwidth,trim={0 0cm 0 0cm},clip]{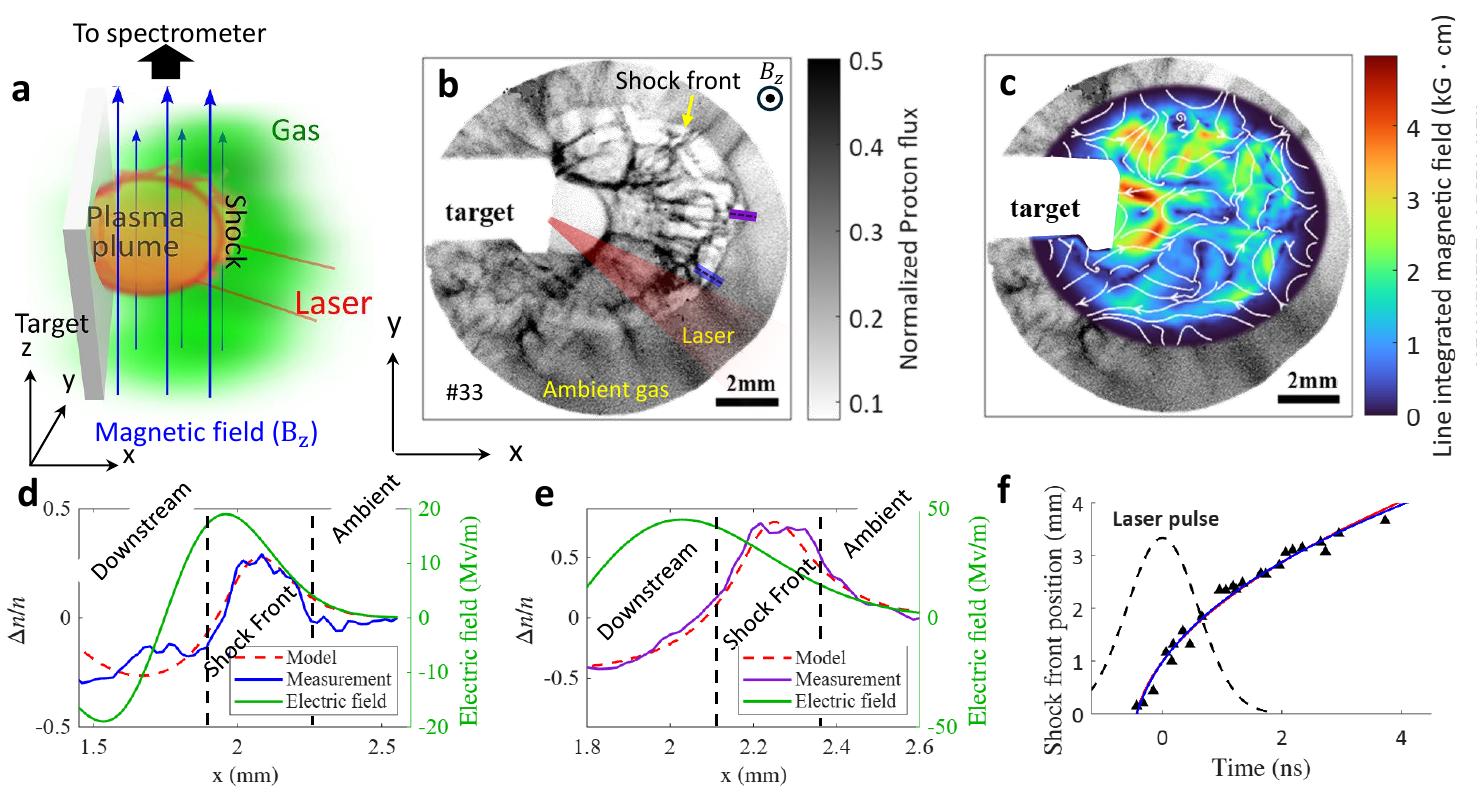}
    \caption{\textbf{Characterization of a single supercritical shock driven by a laser beam in a magnetized gas plasma.}  a. Experimental setup. The ambient gas is hydrogen with, once ionized, an electron number density of $2\times 10^{18}$ cm$^{-3}$ (see Methods). The externally applied magnetic field is set at 10 T. b. Proton radiography image of \textcolor{red}{the expanding plasma and of the shock located at its front.}  
    The protons probe the plasma $2.34$ ns after the temporal peak 
    of the  laser pulse irradiating the target, and the proton energy is $9.7$ MeV. c. Path-integrated map of the magnetic field located behind the shock front (see Section 2.1 of the Supp. Info. for details). 
    \textcolor{red}{
    d. and e. Analysis of the electric field associated with  the shock front (see Section 2.2 of the Supp. Info. for details). Shown are the lineouts of the probing protons density modulations measured at the shock front at two locations  (indicated in panel b. by the blue and purple color lines), together with the simulated modulations (in dashed)  resulting  from the electric field (in green). 
    f. Shock front position as a function of time, as compiled from the proton radiographs, with a  fit by an expansion model  (see Section 1 of the Supp. Info. for details). 
     }
    }
    \label{b_map}
\end{figure*}

\begin{table*}[ht]
\centering
\begin{tabular}{|c|c|c|}
\hline
Quantity                                                & Astrophysical systems & Laboratory          \\ \hline
$\lambda_{\rm mfp}/d_i$                               & $10^7-10^{13}$         & \textcolor{red}{$2\times10^3$}         \\ \hline
${\rho_i}/d_i$                               & $10-10^{3}$         & 14         \\ \hline
Alfvénic Mach number $M_A = v_{sh} / v_A$                              & $\sim 3-100$s      & $\sim 10$          \\ \hline
Magnetic field overshoot $B_{DS}/B_{US}$              & $\sim 2-20$      & $\sim 3-6$                \\ \hline
\end{tabular}%

\caption{\textbf{Key dimensionless parameters of typical astrophysical shocks and those produced in the reported experiment.} The ratio of the proton mean-free-path ($\lambda_{\rm mfp}$) over the shock width, given by the ion inertial length $d_i$, designates the collisionality of the system (\textcolor{red}{see also section 4 of the Supp. Info.}). 
The proton mean-free-path is given by $\lambda_{\rm mfp}=v_{sh,avg}\,\tau_i$, where $v_{sh,avg}$ is the \textcolor{red}{distance-averaged} shock velocity
(\textcolor{red}{see also Fig. S1}) and $\tau_i$ is the ion \textcolor{red}{slowing-down time obtained from the NRL fast-beam formula \cite{NRL_Formulary} ($\lambda_{\rm mfp}\propto v_{sh}^{4}$), evaluated at the measured downstream conditions of the colliding-shock case ($T_i=200$~eV, $n_i=10^{19}~\mathrm{cm}^{-3}$, $\ln\Lambda=10$)}. 
$\rho_i$ is the  Larmor radius of upstream (US) protons having, in the reference frame of the shock, an incoming velocity equal to that of the shock in the laboratory frame. \textcolor{red}{Note that at the shock front, the compressed magnetic field induces small $\rho_i \sim$ 0.26 mm, allowing the protons to perform many gyrations within the available system size.} The Alfvénic Mach number  $M_A$ is calculated with respect to the US values, i.e. it is given by the ratio between  $v_{sh}$  and the Alfvén velocity $v_A \equiv B/\sqrt{\mu_0 n_pm_p}$, where B is the magnetic field strength, $m_p$ the proton mass and $n_p$  the proton plasma density. For the laboratory shock, $T_i$ $\approx$ 200 eV and $n_p$ $\approx 10^{18} \ \rm cm^{-3}$. 
\textcolor{black}{For interplanetary shocks  \cite{
Oliveira2023,
Lindberg2025},  $n_p$ $\approx 20 \ \rm cm^{-3}$, $T_i$ $\approx 10 \ \rm eV$, B $\approx 10 \ \rm nT$ and $M_A$ is in the range of $3-20$. For shocks induced by SN remnants \cite{vink_physics_2020} in the interstellar medium (ISM), we use $T_i$ $\approx 300 \ \rm eV$, $B \approx 1-5 \ \mu$G for the  ISM magnetic field  \cite{Crutcher2012
} and $0.01-10$ cm$^{-3}$ for its density. }
Since SN remnants  have velocities ranging from a few thousands km/s, for young ones \cite{Parizot2006}, to the $1000-5000$ km/s range, for older ones \cite{Martin2018}, we thus have $M_A$ in the several tens-hundreds range.
In galaxy clusters \cite{McNamara2005,Molnar2016}, $T_i$ $\approx$ 2 keV, $n_p$ $\approx 10^{-2} \ \rm cm^{-3}$, and  $v_{sh}$ is in the range $1000-2000$ km/s with $B \approx 0.1-1 \ \mu$G, so we also have $M_A \sim 100$. 
Finally, for lobes and winds of active galactic nuclei \cite{FaucherGigure2012}, their velocity  can be up to $10000$ km/s or more, and in the circumgalactic medium we have $0.01-0.1 \ \mu$G at a density  of $10^{-2}-10^{-3}$ cm$^{-3}$, with $T_i$ $\approx$ 1 keV, which results in  $M_A \sim 100-1000$.
As for the overshoot \cite{
Gedalin2023} 
of the  downstream (DS) magnetic field with respect to the US one, it  is calculated as $B_{DS}/B_{US}$. 
}
\label{tab:parameters}
\end{table*}


We begin by characterizing the individual shocks produced in the experiment (see Methods). 
As illustrated in Fig.~\ref{b_map}a, using a high-power laser, we create a high-velocity plasma piston from an irradiated target. 
As this piston expands into a magnetized, tenuous hydrogen ambient gas, a shock is driven \cite{schaeffer2019direct,yao2021laboratory}. 
To diagnose the shock dynamics, we employ ultra-fast proton radiography \textcolor{red}{and Thomson scattering} (see Methods). 
\textcolor{red}{
The former allows us to locate the shock front via local pile-up of probing protons induced by the shock’s electric field (Figs.~\ref{b_map}d,e). The latter reveals density and temperature jumps that are correlated in space and time (see Supplementary Information). Together, these two diagnostics attest rapid formation of a fast shock within the laser pulse ramp-up phase (Fig.~\ref{b_map}f).}

\begin{figure*}[htp]
    \centering
    \includegraphics[width=1.0\textwidth]{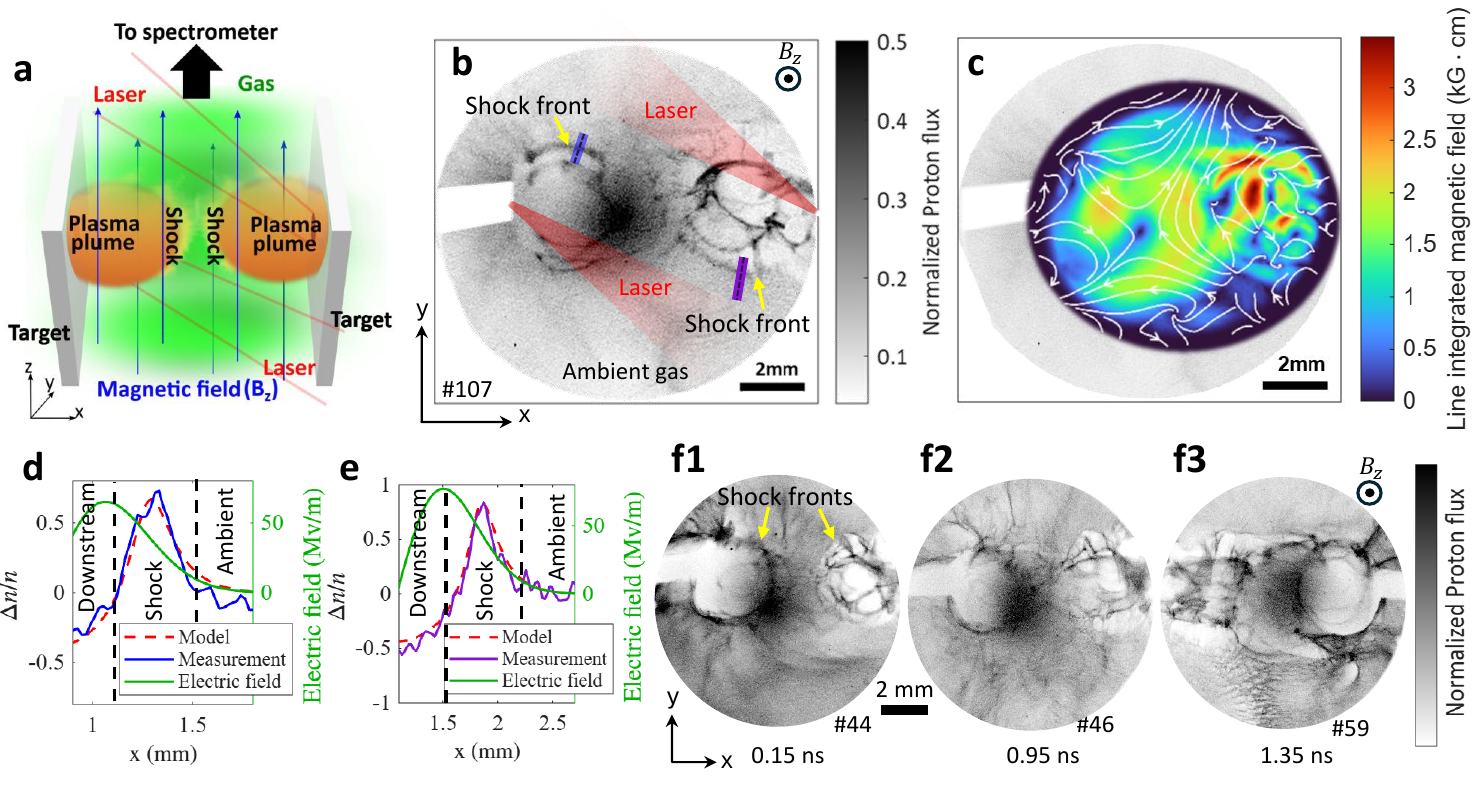}
    \caption{\textbf{Characterization of the collision between two supercritical shock produced in the laboratory experiment.} a. Experimental setup (see Methods). 
    \textcolor{red}{b. Proton radiography image of the interpenetration of  two  shocks and c. Corresponding path-integrated map of the magnetic fields. 
    The protons probe the plasma 1 ns after the temporal peak of the laser pulse irradiating the target, and the proton energy is 9.7 MeV.
    d. and e. show the lineout of the probing protons density at the  two different shock fronts (indicated in b. in blue and purple colors), which are fitted to a model of electric field deflection as detailed in Section 2.2 of the Supp. Info.}
    f1-f3. Proton radiography of the interacting shocks at different times. f1. before (0.15 ns), f2. during (0.95 ns); f3. after (1.35 ns) the collision.  Time $t=0$ refers to the moment when the peak of the laser pulse reaches the target (see Fig.1f). The ambient gas is hydrogen with, once ionized, an electron density of $2\times 10^{18}$  cm$^{-3}$ and the target distance is 4 mm. 
    }
    
    \label{exp_setup}
\end{figure*}

Fig.~\ref{b_map}b presents a snapshot of an \textcolor{red}{evolved} single piston and \textcolor{red}{of} the resulting shock, \textcolor{red}{
while Fig.~\ref{exp_setup}f1 shows the complementary early phase. 
Despite inhomogeneities and filamentary structures in the probing proton beam \cite{Fuchs2003,Metzkes2014,Gde2017}, all images show a sharp interface at the front of the expanding plasma, where protons pile up in a thin layer.
This pile-up is consistent with the expected presence, at the shock front, of a strong electric field. 
Figure~\ref{b_map}d-e and Figure~\ref{exp_setup}d-e show the measured proton pile-up at the interface and quantify the corresponding electric field.
Figure S4 in the Supplementary Information details the temporal evolution of the shock front electric field, which decreases as the shock expands.
The presence of a shock is also supported by the Thomson scattering measurements (see section 3 of the Supp. Info.).} 

\textcolor{red}{Behind the shock front, chaotic structures are observed in the proton radiographs. The energy dependence of these features indicates that magnetic fields cause the induced proton deflections (see Sec. 2.1 of the Supplementary Information).} 
\textcolor{red}{Using the PROBLEM algorithm \cite{bott2017} (see Methods), we obtained path-integrated magnetic field maps, such as that shown in Fig.~\ref{b_map}c (along the $z$-axis, the direction of the applied field). These maps attest to the presence of strong magnetic fields downstream of the shock.}
As anticipated from both astrophysical observations and simulations \cite{balogh2013physics,vink_physics_2020}, these maps allow us to observe a rim of compressed magnetic field right behind the shock front. The compressed magnetic field has a field strength 3 to 6 times higher than in the upstream. This enhancement aligns with expectations for supercritical astrophysical shocks \cite{Mellott1987, Burlaga2008,Gedalin2023,Lindberg2025}, as summarized in Table~\ref{tab:parameters}. Consistent with satellite measurements \cite{Burlaga2008,Johlander2016} and astrophysical simulations \cite{vanMarle2017}, we also observe that the compressed magnetic region exhibits strong corrugation, likely due to instabilities \cite{burgess2007shock}. 

\textcolor{red}{ Fig.~\ref{b_map}f summarizes the position of the shock front, as recorded over multiple shots performed at different times, following the target irradiation by the drive laser (the envelope of which is shown  by the dashed line). From these, we can deduce (see Section 1 of the Supp. Info.) the velocity of the shock, which is observed to decrease over time, as expected. Importantly, up to the first ns of the shock formation and propagation}, the average shock propagation velocity is measured to be $\sim 3000\,\rm km/s$. Given the plasma density and magnetization, this corresponds to an Alfvénic Mach number of $M_A > 10$, similar to shocks observed in diverse astrophysical systems (Table~\ref{tab:parameters}).


\textcolor{red}{Now, we can then symmetrize the system, and, using two targets and two lasers, generate} two counter-facing supercritical shocks. The setup is illustrated in Fig.~\ref{exp_setup}a. We use the same diagnostic technique as in the single-shock case to sample the shock dynamics. The two targets are positioned 4 mm apart (see Methods), ensuring the shocks interact early \textcolor{red}{i.e. after around 1 ns of evolution, when } 
their individual velocities remain high. Fig.~\ref{exp_setup}b shows that proton radiography accurately captures the collision dynamics. \textcolor{red}{The complementary frames shown in Fig.~\ref{exp_setup}f1-f3 show the phases:} before, during, and after the collision -- as well as the subsequent merging of the two structures. \textcolor{red}{Figure~\ref{exp_setup}c details the line-integrated magnetic fields downstream of the shocks around the time of their encounter, further confirming the presence of strong magnetic fields in these regions. These fields will drive strong convective electric fields in the outflows, which in turn will boost the energy of the particles initially energized by the individual shocks, as we will show below.}

The Thomson scattering \textcolor{red}{measurements (see Section 3 of the Supp. Info.) show that even stronger density and temperature jumps are associated with the encounter of the two shocks}, as the shock collision introduces an additional heating source to the plasma \cite{yao2023investigating}. These measured values are used to calculate the ion collisional time used in Table~\ref{tab:parameters}.

\begin{figure}[htp]
    \centering
    \includegraphics[width=0.5\textwidth]{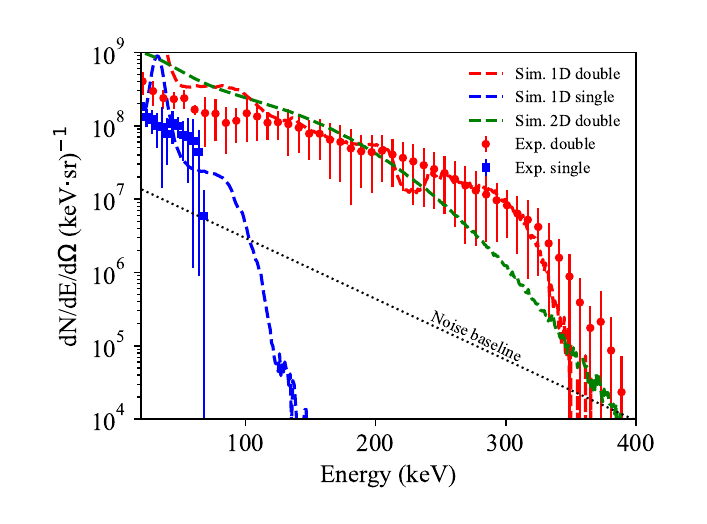}
    \caption{\textbf{Spectra of the protons energized following the shock(s) interaction with the  ambient}. The graphs show the spectra that are measured in the experiment (red dots for the double shock case and blue dots for the single one), 
    and in the simulations (\textcolor{red}{red for the 1D double shock case, blue for the 1D single one, and green for the 2D double one}). 
    The experimental spectra are time-integrated and result from collecting the protons along the z-axis (i.e. the same axis as the applied magnetic field). 
    The simulated spectra are snapshots taken at the end of the simulation \textcolor{red}{around 4 ns}.
    The horizontal axis represents the kinetic energy of the protons, while the vertical axis represents the number of protons per bin of energy (dN/dE), divided by the solid angle (d$\Omega$) subtended by the entrance pinhole of the spectrometer (for the experimental spectra). The single (resp. double) shock spectrum is an average over five (resp. six) shots performed in the same conditions \textcolor{red}{(see Section 5.1 of the Supp. Info. for the individual spectra)}. The dots represent the averaged values, while the error bars correspond to the 1$\sigma$ deviation from the average. 
    The noise level of the diagnostic \textcolor{red}{(inferred by applying the same signal deconvolution procedure used for the spectra to the background on the exposed detector)} is indicated by the black dotted line. Note that the absolute scale in proton numbers applies only to the experimental spectra, the simulated spectrum is arbitrarily scaled to them.}
    \label{exp_spec}
\end{figure}

\textcolor{red}{The encounter of the two shocks is not only observed to induce a stronger density compression and heating of the plasma, but also to provide a significant energy boost to the protons picked up from the ambient hydrogen plasma. This can be seen in their } energy spectra, as collected along the external magnetic field (see Fig.~\ref{b_map}a and Fig.~\ref{exp_setup}a for the arrangement of the spectrometer in the experiment). The proton spectra are shown in Fig.~\ref{exp_spec}. For a single shock, ambient protons (blue) are energized -- likely via shock surfing acceleration \cite{lever2001shock} -- up to a cutoff energy of $\sim 80\,\rm keV$, consistent with our previous results \cite{yao2021laboratory}. 
Integrating the spectrum and assuming isotropic emission over $2\pi\,\rm sr$ leads to  $\sim 3 \times 10^{10}$ ($\sim 5\,\rm nC$) accelerated protons above an energy of 20 keV, with a total kinetic energy of 0.18 mJ. \textcolor{red}{This represents an upper estimate, as it assumes isotropic emission; however, our 2D PIC simulations (detailed below) show no pronounced directional preference in the x-y plane perpendicular to the B field (see also Fig. S13).} From energy conservation (see Section 1 of the Supp. Info.), the energy contained within one shock is estimated to be \textcolor{red}{$\sim 30\,\rm J$}. Given the 4 mm target separation, upstream protons (having, in the reference frame of the shock, an incoming velocity equal to that of the shock in the laboratory frame) undergo $\gtrsim 2$ Larmor gyrations over this distance. From this, we can infer a shock-to-proton energy conversion efficiency of \textcolor{red}{$\sim 0.003\,\%$} per Larmor period. This value aligns well with simulation predictions \cite{caprioli2014simulations1}, further validating the astrophysical relevance of our laboratory shocks.

Remarkably, when two shocks collide (red points, Fig.~\ref{exp_spec}), the proton cutoff energy increases fivefold to $\sim 400\,\rm keV$, while the total energy in accelerated protons rises to $\sim 2.7\,\rm mJ$, corresponding to $\sim 2\times 10^{11}$ protons ($\sim 30\,\rm nC$). This represents an eightfold increase in shock-to-proton energy conversion efficiency compared to the single-shock case. 
\textcolor{red}{We note that we did not observe any energetic protons emerging from the system in the absence of either the ambient gas, or of the external magnetic field driving the shocks. 
The double-shock spectrum exhibits a much flatter shape than expected for shock acceleration processes \cite{Guo2014}, precisely because the ion energy boost arises not from the shocks themselves but from their respective downstream convective electric field -- a mechanism we analyze below using PIC simulations.
}


It is also worth noting that, when the two colliding shocks are subcritical -- as achieved by increasing the standoff distance between the piston-generating targets to 7~mm (see \textcolor{red}{section 5.1 of the Supp. Info.}) -- no further boost in the energy of ambient protons is observed. This result aligns with our earlier experiment, in which two shocks collided $\sim 10\,\rm ns$ after piston initiation, with a magnetosonic Mach number of only 1.1 at the time of collision \cite{fazzini2022particle}, and no proton reenergization was detected.


To \textcolor{red}{shed light onto the mechanism leading to the energy boost }observed in  our experimental results, we have performed kinetic particle-in-cell (PIC) simulations using the SMILEI code \cite{derouillat2018smilei}. The simulations were conducted in 1D3V \textcolor{red}{and 2D3V} geometry (1D\textcolor{red}{/2D} in space and 3D in momentum space). \textcolor{red}{Instead of modeling the piston-plasma interaction, we directly initialized the simulations with experimentally measured shock parameters} (density, temperature, velocity, and magnetic field).

\begin{figure*}[htp]
    \centering
    \includegraphics[width=0.9\textwidth]{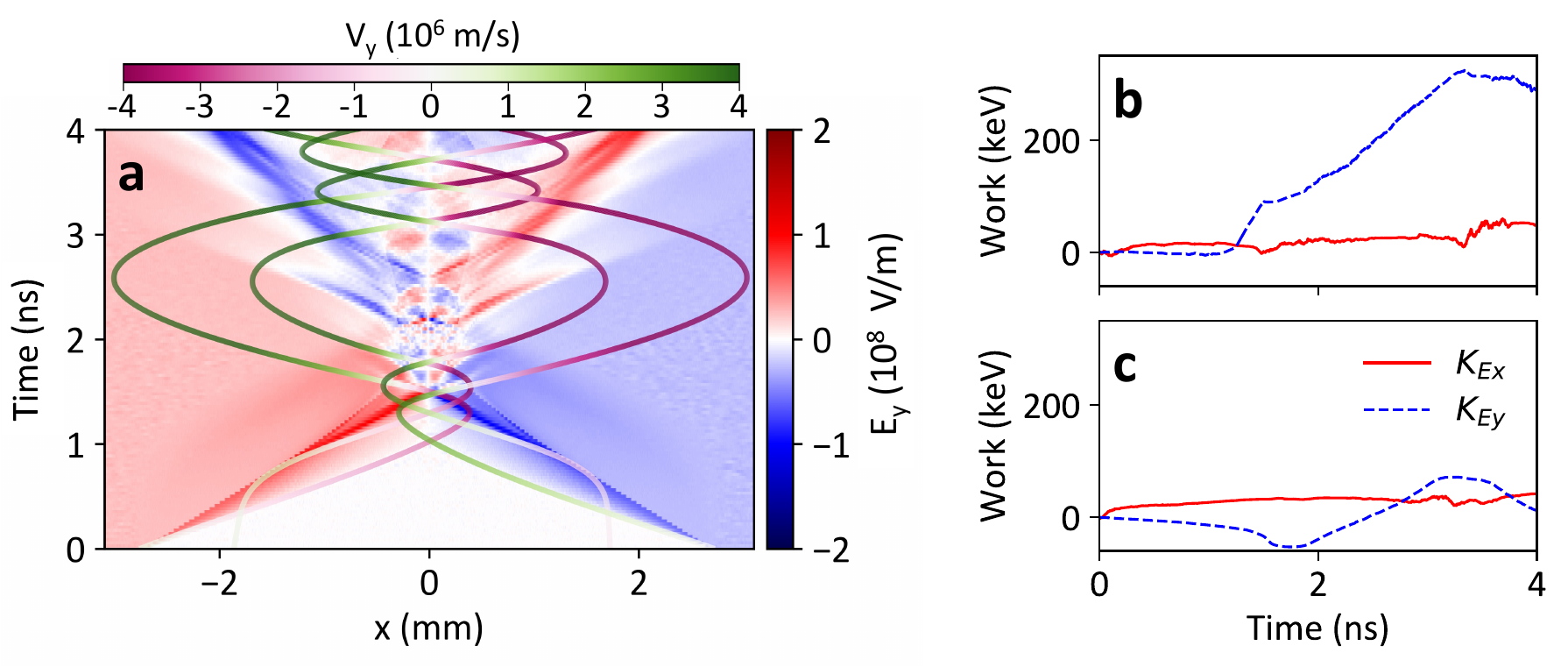}
    \caption{\textbf{Detailed dynamics of the protons in the ambient medium interacting with the colliding shocks, as retrieved from the simulations.}  a. Test particle trajectories with a colormap (on top) representing the transverse velocity ($v_y$), overlaid on the spatiotemporal evolution of the convective electric field ($E_y$), with a colormap on the right. b and c. Temporal evolution of the work performed on a single particle by the electrostatic ($E_x$, red solid) and the convective electric ($E_y$, blue dashed) field in b. the double-shock and c. the single shock case.
    }
    \label{simu}
\end{figure*}

Fig.~\ref{simu}a illustrates the spatiotemporal evolution of the convective electric field $E_y$ in the colliding-shock case. The two shocks form within the first 1~ns of the simulation, and propagate at an initial velocity of $v_{sh} \approx 2750 \pm 250\,\rm km/s$, close to the experimental observation and corresponding to an initial kinetic energy of $E_{sh} \equiv 0.5 m_p v_{sh}^2 \approx 40\,\rm keV$.




The shocks collide at $1.25 \pm 0.25\,\rm ns$ at the center of the simulation box ($x=0$), where the simulated local plasma density increases tenfold compared to its unperturbed value (\textcolor{red}{see Section 6 of the Supp. Info. and Fig. S12}), with their velocities decreasing to $1000 \pm 200\,\rm km/s$.


Particle tracking allows us to identify the energization mechanism at work during the shock collision. Fig.~\ref{simu}a shows four representative particle trajectories overlaid on the field evolution map. Their transverse velocity, $v_y$, is indicated by the color scale at the top. After reflecting off either shock front, these particles gain most of their energy by traveling back and forth in the unperturbed regions of the two flows. There, as they propagate against either flow, they are energized by the local convective electric field, i.e., when $E_y$ is positive (or negative), the particles have their $v_y$ also being positive (or negative) \cite{fazzini2022particle,yao2023investigating}.


This is confirmed in Fig.~\ref{simu}b, where we plot the work performed on a single particle by the convective ($E_y$) and electrostatic ($E_x$) fields. While the latter accounts for the initial modest energy gain ($\sim 10 \,\rm keV$), the convective field largely dominates after $t\approx 1.5 \,\rm ns$, with its work eventually exceeding $300\,\rm keV$ by $t \approx 3 \,\rm ns$. In stark contrast, for a single shock -- where the second colliding flow and its associated motional field are absent -- the energization is significantly weaker. Here, the energy gain from $E_x$ remains comparable, but the contribution from $E_y$ oscillates between gain and loss, likely due to shock surfing (Fig.~\ref{simu}c).

Fig.~\ref{exp_spec} shows that the simulated overall proton spectra for both configurations are in excellent agreement with the observations, with a cutoff energy approaching $10\,E_{sh}$ in the double-shock case. 
\textcolor{red}{Given the good agreement between measured proton spectra and simulations initialized with experimental plasma parameters, convective electric fields downstream of the shocks offer a plausible explanation for the observed proton energy boost, which occurs only in the colliding-shock configuration.}
\textcolor{red}{We note again here that our setup does not involve prescribed turbulence, and that turbulence is not invoked in being here responsible for the observed energy boost resulting from shock interpenetration.}

\textcolor{red}{
Beyond demonstrating the re-acceleration of ions by convective fields downstream of interpenetrating shocks, the platform we have established could, in the future, address the distinct issue of particle acceleration in shock-turbulence interactions. Recent laser-driven experiments have shown that controlled magnetic turbulence can be produced in the laboratory \cite{bott2017,yao_laboratory_2022}. Combining this capability with our ability to generate supercritical shocks would allow us to study their interaction, thereby enabling direct investigation of the fundamental processes \cite{1977DoSSR.234.1306K,1977ICRC...11..132A,bell2013cosmic} that shape cosmic rays -- the universe’s most energetic particles.}

\section*{Methods}

\subsection*{Experimental setup}

The experiment was performed in the target area west (TAW) of  the  Rutherford Appleton laboratory VULCAN laser facility  (U.K.). The experimental setup combines one or two high-power laser produced plasmas, a large-scale background gas, and an externally applied, large-scale, homogeneous and steady (at the spatial and temporal scales of the experiment) magnetic field \cite{albertazzi2013production}.

Each laser pulse, of 1053 nm  wavelength, has 200 J in energy over a  1.4 ns duration in full-width-at-half-maximum (FWHM). The laser temporal shape is Gaussian. Each laser is focused on its respective target with a 200 micron focal spot in diameter with a super Gaussian spatial shape, inducing an on-target peak intensity of about $10^{14}$ W/cm$^2$.

Each target is a Teflon disk (CF$_2$), with a thickness of 2 mm, thicker than the ablation induced by the lasers. The distance between the  laser-irradiated surface of the two targets is either 4 or 7 mm. The setup is immersed in a large-volume
gas, produced by pulsing a gas jet from a nozzle, so that the density is homogeneous in between the targets. The ambient gas is hydrogen, at a low density of $2\times10^{18}$ cm$^{-3}$ for a 7 bars backing pressure.
The externally applied magnetic field is 10 T, and the direction of the magnetic field is along the z-axis (i.e., perpendicular to the shock propagation direction, which is along the x-axis). 

The diagnostics include proton radiography (PR) \cite{schaeffer2023proton}, Thomson scattering (TS) \cite{sheffield2010plasma}, and a magnetic spectrometer for measuring the spectra of the protons.

The probing protons are accelerated from a secondary 15 $\mu m$ Au target using an auxiliary ultra-short 15 ps duration,  250 J energy laser pulse, which is focused to  $10^{19}$ W/cm$^2$ to generate a broadband, divergent proton beam with a maximum energy 15 MeV, through the target normal sheath-acceleration (TNSA) mechanism \cite{schaeffer2023proton}. The proton beam is accelerated along the z-axis, i.e. along the same axis as the external magnetic field.
The protons then propagate through the plasma, after which they are collected by a stack of radio-chromic films (RCF) \cite{schaeffer2023proton} with thin filters between them.  
The different layers in our RCF stack configuration that are appropriate for field reconstruction are sensitive to proton energies of $E= 9.7, 12, 14$~MeV. The dose maps generated by lower energy protons displayed caustics and could not be used  for field reconstruction \cite{bott2017} (see details in Section 2.1 of the Supp. Info.). The distance between the proton source and the plasma was $5.05$~cm, the distance between the plasma and the RCF stack was $7.24$~cm. The resulting magnification was thus $1.69$
. The temporal evolution of single shocks was followed within the temporal interval from -0.5 ns to 3.7 ns with respect to the peak of the laser pulse.  
The \textcolor{red}{analysis of the electric and} magnetic field\textcolor{red}{s that can be inferred from the proton radiographs is discussed in details in Section 2 of the Supp. Info.}

For TS, we also used another auxiliary laser, this time a long-pulse (also 1.4 ns FWHM in duration), frequency-converted to 527 nm, and which was propagated into the plasma along the z-axis, alternatively to the protons used for PR. The Thomson scattered light was collected in the vertical direction, to ensure measurements in the collective regime, and was analyzed by optical spectrometers coupled to streak cameras, as in our previous experiment on single shocks \cite{yao2021laboratory}. \textcolor{red}{A detailed analysis of the TS results is given in Section 3 of the Supp. Info.}

 To measure the proton spectra, we collected the energized ambient protons  along the z-axis (i.e. along the magnetic field axis), alternating with the PR and TS diagnostics. As in our previous experiment on single shocks \cite{yao2021laboratory}, we used a calibrated magnetic spectrometer to resolve the energy of the protons. The protons were detected, in a time-integrated manner, using absolutely calibrated BAS-TR imaging plates \cite{martin2022absolute} equipped with filters to ensure that ions other than protons did not reach the detector. \textcolor{red}{
 We placed a 1-mm-diameter pinhole at the entrance of the spectrometer,  corresponding to an acceptance solid angle of $3.5\times 10^{-5}$ sr. Since the particle energy was relatively low, and thus strongly dispersed, the pinhole size did not impair the energy resolution. The pinhole also caused the spectrometer to act like a pinhole camera. By analyzing the proton trace width, we determined that the signal originated from a 7 mm-wide region within the plasma. However, the 1 mm pinhole introduced some smearing, meaning we could not pinpoint the protons' exact origin. Instead, they appeared to come from the entire plasma, though the source did not fill the full 1 cm coil aperture. This is expected, as the protons were not highly laminar. Additionally, the fringe field of the guide field likely further distorted their trajectories toward the spectrometer.
 }


\subsection*{Simulation setup}

The shock collision is modeled with the fully kinetic Particle-In-Cell (PIC) code SMILEI \cite{derouillat2018smilei} in one-dimensional (1D) \textcolor{red}{and 2D} geometry, as in our previous work \cite{yao2021laboratory, yao2022detailed}. 
\textcolor{red}{In 1D}, the simulation box is initialized to be $L_x = 6144 d_e = 33$ mm, with the spatial resolution $d_x = 0.2 d_e = 1.1 \ \mu$m, with the electron inertial length $d_e = c / \omega_{pe} = 5.3 \ \mu$m, and the electron plasma frequency  $\omega_{pe} = (n_{e0} q_e^2 / m_e  \epsilon_0)^{1/2} = 5.6 \times 10^{13}$ s$^{-1}$. Here, the initial ambient electron number density is $n_{e0} = 1.0 \times 10^{18}$ cm$^{-3}$; $c$, $m_e$, $q_e$ and $\epsilon_0$ are the speed of light, the electron mass, elementary charge, and the permittivity of free space, respectively. The simulation box consists of 30720 cells. For each cell, we put 2048 particles for each species. The simulation lasts for $2.5 \times 10^5 \omega_{pe}^{-1} \sim 4.0$ ns. 
\textcolor{red}{In 2D, due to the limitation of computational cost, we use $L_x = 22 \mathrm{~mm}$, $L_y = 3 \mathrm{~mm}$, and 16 particles per cell; while keeping $d_y = d_x = 0.2 d_e$ (like in the 1D case).} 

For the boundary conditions, the electromagnetic fields are absorbed, while particles are removed.
\textcolor{red}{While in the 2D simulations, periodic boundary conditions are applied to both fields and particles along the transverse dimension.}
Note that enough room is left between the boundary and the plasmas, so that the boundary conditions do not affect the dynamics of the interpenetrating shocks.
The ambient plasma lies in the middle of the simulation box, having a size 
\textcolor{red}{ranging from 3.6 to 5.3 mm, similar to}
those in the experiments. We set two shocked plasmas from the left/right sides, each having a size of about 13 mm \textcolor{red}{(in 1D) and 9 mm (in 2D)}, a density of $2n_{e0}$, and a drifting velocity of $v_0 = 2500$ km/s towards the middle. Note that in our simulation, the ion species is proton with its real mass ($m_p / m_e = 1836$), and the shock width is initialized as one ion inertial length $d_i \approx 200 \ \mu$m in between the ambient plasma and the shocked plasma.

The initial temperature of the ambient plasma is $T_{e0} = T_{i0} = 100$ eV, as in the experiment. 
The initial temperature of the shocked plasma for the double shock case is $T_{e,DS} = T_{i,DS} = 250$ eV, and that for the single shock case is $T_{e,SS} = T_{i,SS} = 190$ eV. The shocked plasma temperature in single shock case is a bit lower than that of the double shock one, as verified by our experimental measurements (see main text).

The externally applied magnetic field is homogeneously distributed along the simulation box. The magnetic field direction is along the $z$-axis, with a strength of 10 T, i.e., $B_z = 10$ T (which corresponds to $\omega_{ce} / \omega_{pe} = 0.03$, where $\omega_{ce} = q_e B / m_e$). 
Note that during the shock collision, the magnetic field in the downstream around the shock has a compression ratio over 10. Thus, for the protons drifting along the shock, with an initial velocity of $v_{x0} = 2500$ km/s, their Larmor radius is $r_{Li} = 0.26$ mm and gyro-period is $\tau_{Li} = 0.6$ ns.


\subsection*{Acknowledgments}
The authors would like to thank the excellent laser, engineering and target fabrication teams of the Central Laser facility (U.K.)  for their expert support and dedication. This work was supported by funding from the European Research Council (ERC) under the European Union Horizon 2020 research and innovation program (Grant Agreement No. 787539, JF).  We acknowledge support by  the IMPULSE project by the European Union Framework Program for Research and Innovation Horizon 2020 (grant agreement no. 871161) , and EPRSC (grant EP/P010059/1). This research was also supported in part by grant NSF PHY-2309135 to the Kavli Institute for Theoretical Physics (KITP).
The computational resources of this work were supported by the HPC resources of MesoPSL financed by the Region Ile de France and the project EquipMeso (reference ANR-10-EQPX29-01) of the programme Investissements d’Avenir supervised by the Agence Nationale pour la Recherche. It was also supported by the National Sciences and Engineering Research Council of Canada (NSERC) and Compute Canada (Job: pve-323-ac, PA), and by NASA (grant 80NSSC24K0173) and NSF (grant AST-2510951).

\subsection*{Author Contributions Statement}

J.F. conceived the project. W.Y., H.A., M.C., S.N.C., R.L., P.M., J.B., T.W. and J.F. performed the experiments, with support from P.A., and M.B. W.Y., I.C., A.F.A.B., S.N.C., and J.F. analyzed the data. W.Y. and P.S-G. performed and analyzed the SMILEI simulations, with discussions with A.C., M.F., T.S., L.G., E.d.H. and J.F. The bulk of the paper was written by W.Y., S.N.C., I.C. and J.F., with inputs from L.G., S.O., M.M., A.M. and D.C. All authors commented and revised the paper.

\subsection*{Materials and correspondence}
Correspondence and material requests should be addressed to
julien.fuchs@polytechnique.edu.

\subsection*{Competing Interests Statement}
The authors declare no competing interests.

\textbf{\textit{Data availability}}
All data needed to evaluate the conclusions in the paper are present in the paper.
Experimental data and simulations are respectively archived on servers at LULI and LERMA laboratories and are available from the corresponding author upon reasonable request.

\textbf{\textit{Code availability}}
The code used to generate Fig.~\ref{exp_spec} and Fig.~\ref{simu}
is SMILEI, which is detailed in the Methods section.

\bibliographystyle{naturemag}
\bibliography{main}

\end{document}


\title{Supplementary Information: Efficient ion re-acceleration in laboratory-produced interpenetrating collisionless shocks}

\author[1,2]{W. Yao}
\author[1]{I. Cohen}
\author[1]{P. Suarez Gerona}
\author[3]{H. Ahmed}
\author[4]{A.F.A. Bott}
\author[5]{S. N. Chen}
\author[3]{M. Cook}
\author[1,6]{R. Leli\`evre}
\author[7]{P. Martin}
\author[1,8,9]{T. Waltenspiel}
\author[9]{P. Antici}
\author[10]{J. B\'eard}
\author[7]{M. Borghesi}
\author[11,12]{D. Caprioli}
\author[2]{A. Ciardi}
\author[8]{E. d'Humi\`eres}
\author[8]{M. François}
\author[13,14]{L. Gremillet}
\author[15]{A. Marcowith}
\author[16]{M. Miceli}
\author[1,2,17]{T. Seebaruth}
\author[18]{S. Orlando}
\author[1]{J. Fuchs}

\affil[1]{LULI - CNRS, CEA, UPMC Univ Paris 06 : Sorbonne Universit\'e, Ecole Polytechnique, Institut Polytechnique de Paris - F-91128 Palaiseau cedex, France}
\affil[2]{Sorbonne Université, Observatoire de Paris, Université PSL, Laboratoire d'étude de l'Univers et des phénomènes eXtrêmes, LUX, CNRS, F-75005 Meudon, France}
\affil[3]{Central Laser Facility, STFC Rutherford Appleton Laboratory, Oxfordshire OX11 0QX, United Kingdom}
\affil[4]{Department of Physics, University of Oxford, Parks Road, Oxford OX1 3PU, United Kingdom}
\affil[5]{ELI-NP, "Horia Hulubei" National Institute for Physics and Nuclear Engineering, 30 Reactorului Street, RO-077125, Bucharest-Magurele, Romania}
\affil[6]{Laboratoire de micro-irradiation, de métrologie et de dosimétrie des neutrons, PSE-Santé/SDOS, IRSN, 13115 Saint-Paul-Lez-Durance, France}
\affil[7]{Center for Plasma Physics, School of Mathematics and Physics, Queen's University Belfast, Belfast BT7 1NN, United Kingdom}
\affil[8]{University of Bordeaux, Centre Lasers Intenses et Applications, CNRS, CEA, UMR 5107, F-33405 Talence, France}
\affil[9]{INRS-EMT, 1650 boul, Lionel-Boulet, Varennes, QC, J3X 1S2, Canada}
\affil[10]{LNCMI, UPR 3228, CNRS-UGA-UPS-INSA, Toulouse 31400, France}
\affil[11]{Department of Astronomy \& Astrophysics, University of Chicago, Chicago, Illinois 60637, USA}
\affil[12]{Enrico Fermi Institute, The University of Chicago, Chicago, Illinois 60637, USA}
\affil[13]{CEA, DAM, DIF, F-91297 Arpajon, France}
\affil[14]{Université Paris-Saclay, CEA, LMCE, F-91680 Bruyères-le-Châtel, France}
\affil[15]{Laboratoire Univers et Particules de Montpellier CNRS/Université de Montpellier, Place E. Bataillon, 34095 Montpellier, France}
\affil[16]{University of Palermo, Department of Physics and Chemistry, Palermo, Italy}
\affil[17]{Laboratoire de Physique des Plasmas (LPP), CNRS, Observatoire de Paris, Sorbonne Université, Université Paris-Saclay, École polytechnique, Institut Polytechnique de
Paris, F-91120 Palaiseau, France}
\affil[18]{INAF–Osservatorio Astronomico di Palermo, Palermo, Italy}

\date{\today}

\maketitle

\section{Characterization of the shock dynamics}

\begin{figure}[htp]
    \centering
    \includegraphics[width=0.47\textwidth]{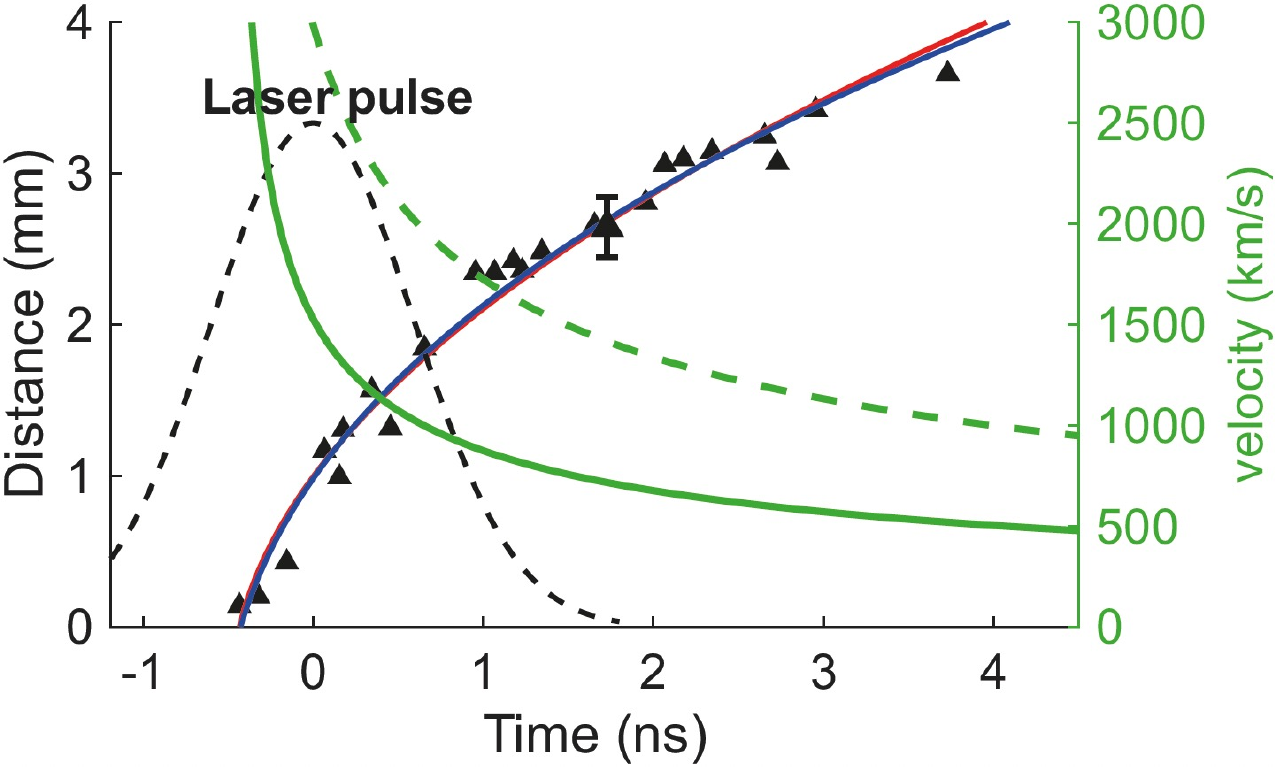}
    \caption{\textbf{Measured single shock front position as a function of time}, overlaid with a fit from shock expansion models (see text for details). \textcolor{red}{The error bar shown on one of the signal points of the measurement is the same for all points and is not shown on all points to ease visualization. The green line is the shock instantaneous velocity inferred from the model, whereas the dashed green line is the averaged velocity from the shock initiation up to its current position. Distance 0 is set at the surface of the laser-irradiated target. Time 0 corresponds to the peak of the irradiating laser pulse (see Fig.1 of the main text for the laser pulse temporal shape).}
    }
    \label{velocities}
\end{figure}

We measured the dynamics of the shock induced by the plasma piston expanding into the magnetized gas. This was done by performing a series of proton radiographic images (see Methods), such as the one shown in Fig. 1 of the main text. From these, we could measure the evolution of the shock front position a a function of time. The result is shown in Fig.~\ref{velocities}. 

The shock  dynamics is measured, from these proton radiography images, in a time window of $-0.43$ ns to $4$ ns with respect to the peak of the laser pulse (which corresponds to t=0). 
We can compare these data with a simple model of the expansion of the shock. In this model, the shock is formed in an initial impulse conferred to it by the piston, knowing that the piston expansion itself is not sustained over time, due to the impulsive nature of the laser energy deposition on the target. In this frame, we thus assume that once the shock is formed, its overall energy is conserved as it expands. Thus, as the volume occupied by the shock increases, its velocity will decrease, as observed.  
In our model, we describe the shock as the volume contained between two spheres which are separated by the shock width (we measure it on the images to be $\Delta=0.16$ mm, which is in good agreement with the expected value given by the ion inertial length \cite{balogh2013physics} $d_i=0.23$ mm). The shock width is assumed to be constant during the shock evolution, consistently with the observations based on the proton radiographs.
The volume occupied by the shock ($V(t)$) is then given by:
\begin{align*}
V(t)=S_{out}-S_{in}
\end{align*}
where
\begin{align*}
S_{out} &= \frac{4\pi R(t)^{3}}{3} ,\\
 S_{in} &= \frac{4\pi (R(t)-\Delta)^{3}}{3} \\
        &\approx \frac{4\pi}{3}\!\left(1-\frac{3\Delta}{R(t)}\right)R^{3} \\
        &= S_{out} - 4\pi\,\Delta\,R(t)^{2},\\
\end{align*}
thus we have
\begin{align*} 
V(t)\simeq 4\pi\,\Delta\,R^{2}(t).
\end{align*}
where $R(t)$ is the shock front position, $\Delta$ is the  shock width, $S_{in}$ is the inner surface area of the shock layer, $S_{out}$ is the outer surface area of the shock layer, and $V(t)$ is the volume occupied by the shock layer. 

The total mass and energy within the shock are given by:
\begin{align*}
M(t) &= m_p\,n\,V(t),\\
E(t) &= \tfrac12\,M(t)\,v^{2}(t)
      = \tfrac12\,m_p\,n\,V(t)\,v^{2}(t).
\end{align*}
where $m_p$ is the proton mass and $n$ is the density of the plasma in the shock layer, \textcolor{red}{inferred from the Thomson scattering measurements (see Table ~\ref{tab:TS_parameters}). }

We assume now that the temporal evolution of $R(t)$ is given by a power law and derive from it the shock position and velocity.
\begin{align*}
R(t) &= R_0\!\left(\frac{t}{t_0}\right)^{\alpha},\\
v(t) &= \dot R(t)=\alpha\frac{R_0}{t_0^{\alpha}}\,t^{\alpha-1}.
\end{align*}

We can assume that the energy of the shock, $E(t)$, is conserved over time. The shock energy is given by:
\begin{align*}
E(t) &=\frac12\,m_p\,n\bigl[4\pi\,\Delta\,R^{2}(t)\bigr]\,v^{2}(t)\\
     &=\frac12\,m_p\,n\left[4\pi\,\Delta\,R_0^{4}\alpha^2\frac{t^{4\alpha-2}}{t_0^{4\alpha}}\right] \\
     &\propto t^{4\alpha-2}.
\end{align*}
Energy conservation thus dictates a value of $\alpha=0.5$. Therefore, the energy of shock is given by:
\begin{align*}
E
&=\frac12\pi\,m_p\,n\,\Delta \frac{R_0^{4}}{t_0^{2}}
.
\end{align*}

\textcolor{red}{
As shown in Fig.~\ref{velocities}, the model (red line) closely agrees with the experimental data. For comparison, a classical Sedov-Taylor expansion \cite{Sanz2016}, which yields the scaling $R(t) \propto t^{2/5}$, produces an identical trajectory during this timescale (blue line).
} 
The result of the fit indicates that the shock forms in the rising edge of the laser pulse, at $t_{int}=-0.49$ ns, with a scaling parameter of $R_0/t_0^{0.5}=2.13$~$\rm mm/ns^{0.5}$, and that the shock front appears at a distance of \textcolor{red}{$d_0=-0.49$} mm from the target position (see Fig.~\ref{velocities}). 

Further, the model allows us to estimate the shock velocity at early times. The average velocity of the shock from the time it is initiated up to $t=0$ ns (i.e., to the peak of the laser pulse) is $2946$ km/s. To account for the fact that the velocity decreases as the shock expands until it will reach the counter-propagating shock (when we generate the two, as reported in the main text), we initialize the PIC simulations detailed in the main paper with a velocity of $2500$ km/s, but we verified that we observe the overall same dynamics of the particle accelerated by the single or double shocks when we initiate the simulations with velocities of $2000-3000$ km/s.

With the above model fitting the data, we can also estimate the energy contained within the shock as \textcolor{red}{30.2 J}. This corresponds to a conversion efficiency of about \textcolor{red}{15.1$\%$} from the laser energy of 200 J.


\section{Inference of \textcolor{red}{electro}magnetic fields  from proton radiographic images}
\subsection{\textcolor{red}{Numerical evaluation of field properties}}

\textcolor{red}{Both E- and B-fields can contribute to the observed proton dose modulations in the proton radiographs. The energy dependence of the radiography signals shows that these fields act primarily in distinct regions: E-fields dominate at the shock’s leading edge, while B-fields dominate within the shocked plasma, where the energy-boost mechanism operates.
}

\begin{figure}[htp]
    \centering
    \includegraphics[width=0.45\textwidth]{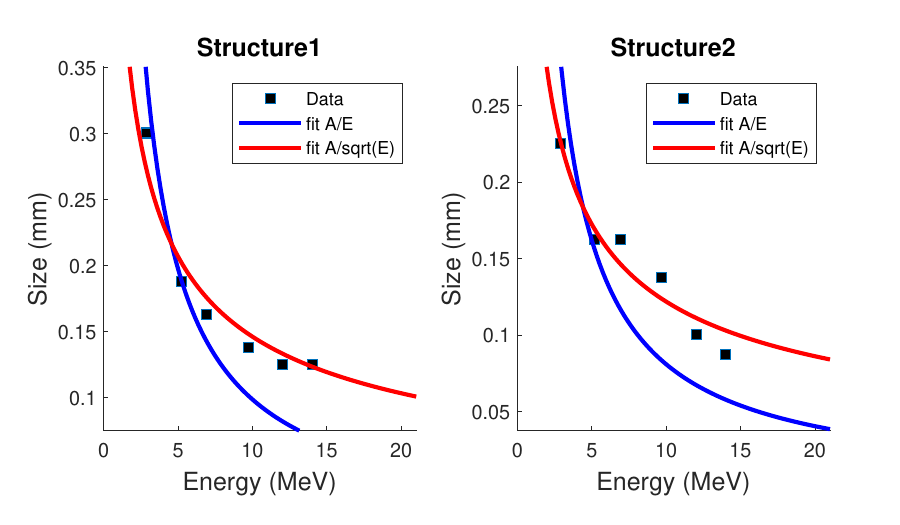}
    \caption{\textbf{Scaling of proton radiography filaments from different RCF layers}. The size of two filaments within the shock front, from the same shot, are analyzed at different proton probing energies. The data is fitted to deflection induced either by a magnetic force or an electric one. }
    \label{EnergyScaling}
\end{figure}

\textcolor{red}{The E-fields at the shock’s outer interface with the ambient medium -- which cause proton pile-up at that interface -- are analyzed in Section 2.2. To infer the magnetic nature of the fields inducing proton deflections behind the shock front, we measured the size of a given filament structure within the shocked plasma.} This measurement was performed on different RCF layers of the same shot, identifying the same filament structure. We assume that the displacement of the protons is equal to the thickness of the filament. The different scaling of the displacement due to the magnetic Lorentz force ($1/\sqrt{E}$) and to the electric one ($1/E$) allows us to verify if the size of the filaments matches one of these scalings \cite{PhysRevE.110.L033201}. These fits are shown in Fig.~\ref{EnergyScaling}, which clearly show that these deflections are due to magnetic forces.

As detailed in the Methods, we then used the PROBLEM code to infer the magnetic field \textcolor{red}{
 in the plasma behind the shock front } from the raw proton radiographic images.
 \textcolor{red}{
 To infer the incident proton beam distribution, we apply a Gaussian blur filter to each image \cite{bott2017}. Doing this removes all the fine gradients induced by the probed fields. We have assessed that using a filter size (the “plasma length parameter”) of 3 mm allows the inferred background proton beam to be as close as possible to the proton flux distributions obtained in the absence of probed plasma}. The use of PROBLEM is justified given that the contrast regime is in the linear regime \cite{schaeffer2023proton}, which can be evaluated using the $\mu$-parameter \cite{bott2017}. The $\mu$-parameter is the ratio of the deflections undergone by the protons to the characteristic length of the plasma. It is defined as: $\mu = \frac{\xi}{M l_B}$, where $\xi$ is the proton lateral displacement in the detector plane, $M$ is the magnification, and $l_B$ is the scale length of the magnetic field. If $\mu$ > 1, the proton trajectories would self-intersect. In this case, the system is non-injective and several magnetic field configurations can produce the same proton radiography maps, preventing the utilization of the field reconstruction. In the data presented here, we estimate that $\mu \lesssim 0.4$, meaning that we are in the linear regime and can perform field reconstruction.

\begin{figure}[htp]
    \centering
    \includegraphics[width=0.45\textwidth,trim={4cm 7cm 4cm 6.9cm},clip]{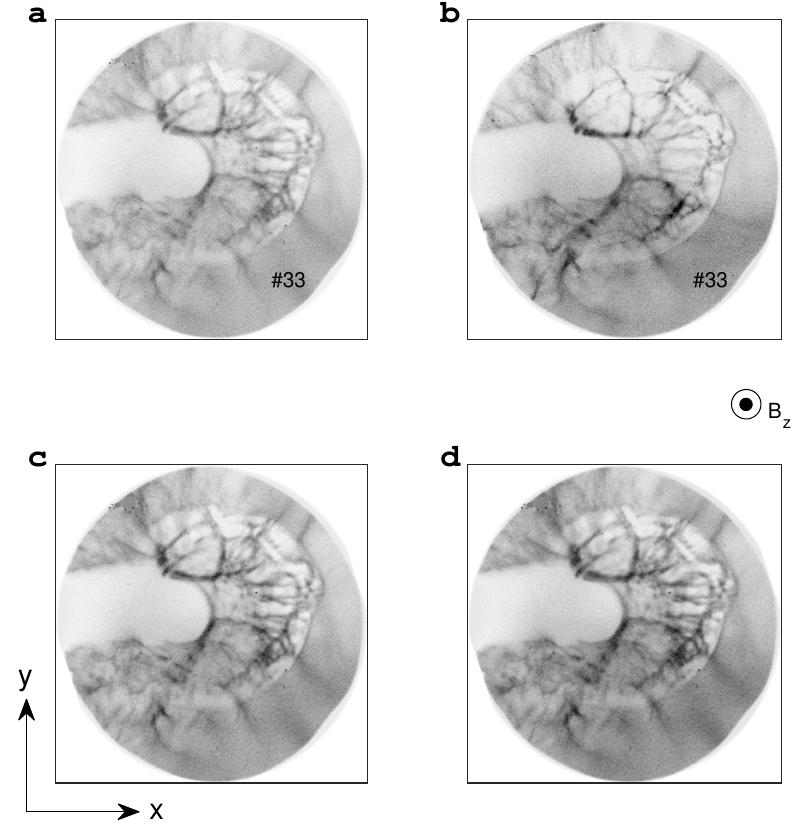}
    \caption{\textbf{Measured proton flux maps and predicted flux maps.} a-b. measured proton flux with energies $E=9.7, 12$ MeV corresponding to probing times of $t=2.34, 2.18$ ns, respectively. c-d. predicted flux maps obtained from the magnetic field retrieved from the dose map shown in panel a, which is shown in Fig.~2 of the main text, 
    for energies of  $E=9.7, 12$ MeV, respectively.
    }
    \label{pre_flux}
\end{figure}

To further verify that the reconstructed fields \textcolor{red}{
 in the plasma behind the shock front } are magnetic, instead of electric, in origin, we test the field reconstruction algorithm by generating artificial proton flux images, with the deflections of simulated protons determined by the reconstructed path-integrated field. 
In this test, we analyze the magnetic field from a specific probing energy and compare the synthetic images of different proton energies to the experimental ones \cite{Tzeferacos2018}.
In Fig.~\ref{pre_flux}a-b we show multiple measured proton radiographic images of a single shock expansion. These radiographic images originate from a single shot, and correspond to the plasma being probed by protons of different energies, as produced by the auxiliary short-pulse laser source \cite{schaeffer2023proton}. We can resolve and separate them due to the arrangement of the detector that uses a stack of radio-chromic films (RCF), each of which is sensitive to a particular range of proton energies \cite{schaeffer2023proton}. Here, we show the radiographic images corresponding to probing protons energies of $E=9.7, 12$ MeV, respectively, which correspond to probing times of $t=2.34, 2.18$ ns. We compare those dose maps to the dose maps shown in Fig.~\ref{pre_flux}c-d, which are predicted dose maps calculated using the PROBLEM code \cite{Tzeferacos2018} 
extracted at $t=2.34$ ns from the dose map of Fig.~\ref{pre_flux}a with energies of $E=9.7, 12$ MeV, respectively. The similarities in the structure sizes and proton dose compression assert that the fields are magnetic, whereas differences are most likely due to the intrinsic temporal evolution of the structures.

Finally, we also considered separating the contribution of the electric and magnetic deflections in the same manner as proposed in Ref.~\citenum{10.1063/5.0033834}. In that analysis, the electric and magnetic deflections are extracted from two RCF films using probing protons of different energies. We found that this analysis yield globally consistent results with those of the field retrieval performed with the PROBLEM code \textcolor{red}{in the downstream plasma}, but that using RCF layers having close-by energies leads to high noise, and using RCF layers with a large energy separation is associated with the field having evolved in time between the frames, and hence to errors in the reconstruction. 

\subsection{\textcolor{red}{Electric field at the shock front}}

\textcolor{red}{While the methodology  reconstruction detailed above allows us to infer the magnetic-field in the downstream plasma, behind the shock front, an electric field is also present at the leading edge of the expansion,  as discussed in the main text and as characterised in detail in our earlier experiments conducted on single supre-critical shock characterization\cite{yao2022detailed}. The signature of this E-field is a sharp proton-dose accumulation followed by a sharp proton-dose depletion at the outer boundary of the shock that is clearly visible in the radiographs (see Fig. 1b and Fig. 2b of the main text), with the same morphology as the one we previously associated with the ambipolar E-field at the shock layer \cite{yao2022detailed}. The electric field along the shock front is given by: $E=\sqrt{2e}E_0\frac{x-g(z)}{L}\exp{\frac{(x-g(z))^2}{L^2}}$
where $E_0$ is the maximum field amplitude and $L$ represents the width of the region affected by the
electric field, z is the axis along which the proton beam propagates. $g(z) = -R/2 +\sqrt{R^2+z^2}$ represents the shift of $E$ along $x$ to take into account the hemispherical geometry in the xz-plane, which is derived from a circumference of radius R centered in $(x,z)=(-R/2, 0)$.}

\textcolor{red}{The amplitude of this E-field has been extracted at different locations and temporal snapshots, following the same methodology as in Ref.\cite{yao2022detailed}: the proton-dose modulation across the shock front is fitted with the pattern produced by a hemispherical bipolar E-field threading the shock layer. Corresponding fits and reconstructed $E$-field profiles are shown in Fig. 1d and 1e (single-shock case) and Fig. 2d and 2e (double-shock case). In this experiment, at early time, shock fronts yield peak amplitudes in the range $E \sim 64\pm14~\mathrm{MV/m}$, more than an order of magnitude stronger than the 4 MV/m value measured in Ref.\cite{yao2022detailed}. This is consistent with the higher Mach number of the present shocks. }

\begin{figure}
    \centering
    \includegraphics[width=0.4\textwidth]{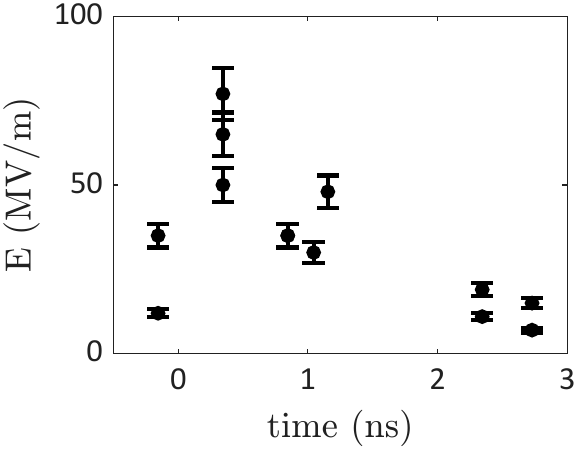}
    \caption{\textcolor{red}{Peak value of electric field at the shock front measured at different positions from different proton radiographs performed at different times with respect to t=0 (which is the peak of the laser pulses driving the plasma piston from the solid targets). The peak value of the E-field shown here is extracted from fitting the proton dose modulation recorded at the shock front, as shown in Fig.1 and 2.}
    }
    \label{fig:Efield_vs_time}
\end{figure}

\textcolor{red}{Fig.~\ref{fig:Efield_vs_time} summarises this analysis across different radiographs probing at various times. The shock front takes $\sim$ 0.4 ns after the peak of the drive laser  to fully develop, reaching a peak amplitude $E_{peak} \sim 64 \pm$ 14 MV/m, after which it decreases monotonically as the shock expands. }

\section{\textcolor{red}{Thomson scattering characterization of the plasma density and temperature jumps induced by the shocks}}

\subsection{{\textcolor{red}{Thomson scattering laser probe}}}
\begin{figure}
    \centering
    \includegraphics[width=0.4\textwidth]{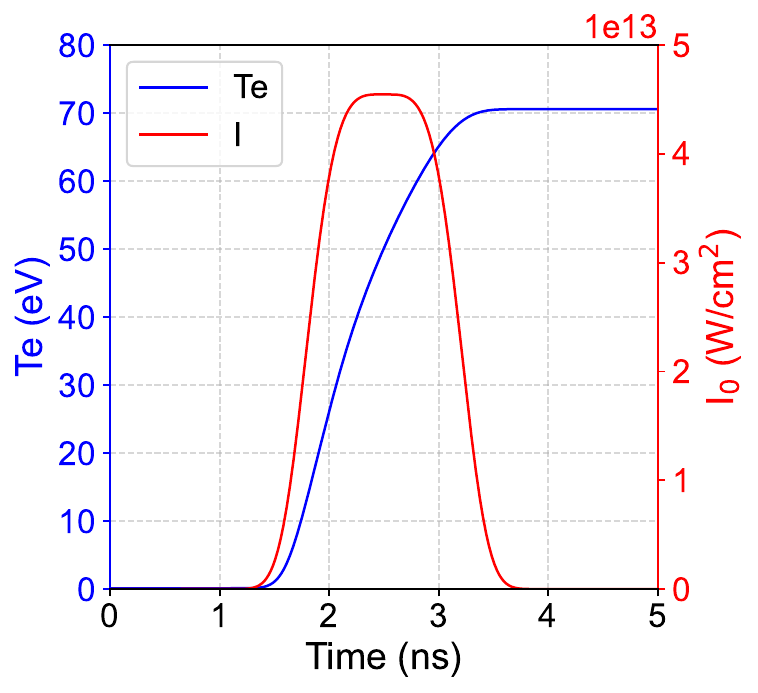}
    \caption{\textcolor{red}{Inverse Bremsstrahlung calculation of the heating of the $2\times 10^{18} \rm ~cm^{-3}$ density ambient H plasma, as induced by a laser having the parameters of the TS laser probe beam. Shown in the temporal evolution of the laser intensity and the induced increase in the electron temperature (in red, $\text{T}_e$) of the plasma (in blue).}
    }
    \label{fig:TS_heating}
\end{figure}

\textcolor{red}{The Thomson scattering (TS) probe laser beam, having a 1.4 ns FWHM  duration, a 527 nm wavelength and 20 J of energy.
The TS probe beam was focused over a wide area, $200~\mathrm{\mu m}$ in diameter, to ensure sampling of a significant plasma volume and to keep the probe laser intensity low enough to avoid strong plasma heating. As shown in Fig.~\ref{fig:TS_heating}, inverse Bremsstrahlung estimates \cite{Yao2023Dynamics} indicate that the TS probe itself would heat the plasma to an electron temperature of 70 eV. Thus, the TS probe beam, along with radiation from the ablated solid targets, likely contributes to the pre-shock heating at the center of the system. However, the 3-4 times jump in electron temperature following the passage of the shock (to $\sim 200 \rm~eV$ for a single shock and $\sim 250 \rm~eV$ for a double shock, see Fig.~\ref{fig:TSe}) cannot be attributed to the TS probe, but is clearly due to shock-induced heating.} 

\subsection{{\textcolor{red}{Thomson scattering measurements}}}
\textcolor{red}{We fielded both TS on the electron waves (TSe) and on the ion waves (TSi) in the plasma.
The TSi and TSe scattered light waves were analysed by means of two different spectrometers, set to different dispersions, and which were coupled to two streak cameras (Hamamatsu for TSe and TitanLabs for TSi; both were equipped with an S-20 photocathode (for sensitivity in the visible part of the spectrum and with a typical 30 ps temporal resolution). The central openings of both streak cameras and spectrometers were imaging  the same location in the plasma (located 2 mm away from each of the solid target surfaces, i.e. the center of the system) within the magnetic field coil, to ensure that the value of the electron density obtained from the TSe analysis corresponds to the same region of plasma that was observed in the corresponding TSi spectrum.
Thin strips of black filters were positioned at the entrance slits of both streak cameras to block the Rayleigh scattered light at the wavelength of the probing laser for both TSi and TSe. 
 The probing volume in the plasma is determined by the intersection between the TS laser probe envelope and the projection of the TS spectrometers and streak cameras. The scattering volumes sampled by the instruments were as follows:
$200~\mathrm{\mu m}$ along the $x$ axis, $200~\mathrm{\mu m}$ along the $y$ axis, and $70~\mathrm{\mu m}$ along the $z$ axis for $\text{TS}_i$; $100~\mathrm{\mu m}$ along the $x$ axis, $200~\mathrm{\mu m}$ along the $y$ axis, and $66~\mathrm{\mu m}$ along the $z$ axis for $\text{TS}_e$.}

\textcolor{red}{We performed shots varying the delay between the TS probe laser and the main lasers irradiating the targets, from which the plasma plumes inducing the shocks were driven. This way, we could characterize the plasma before and after the shocks reached the center of the system.}

\begin{figure*}
    \centering
    \includegraphics[width=0.7\linewidth]{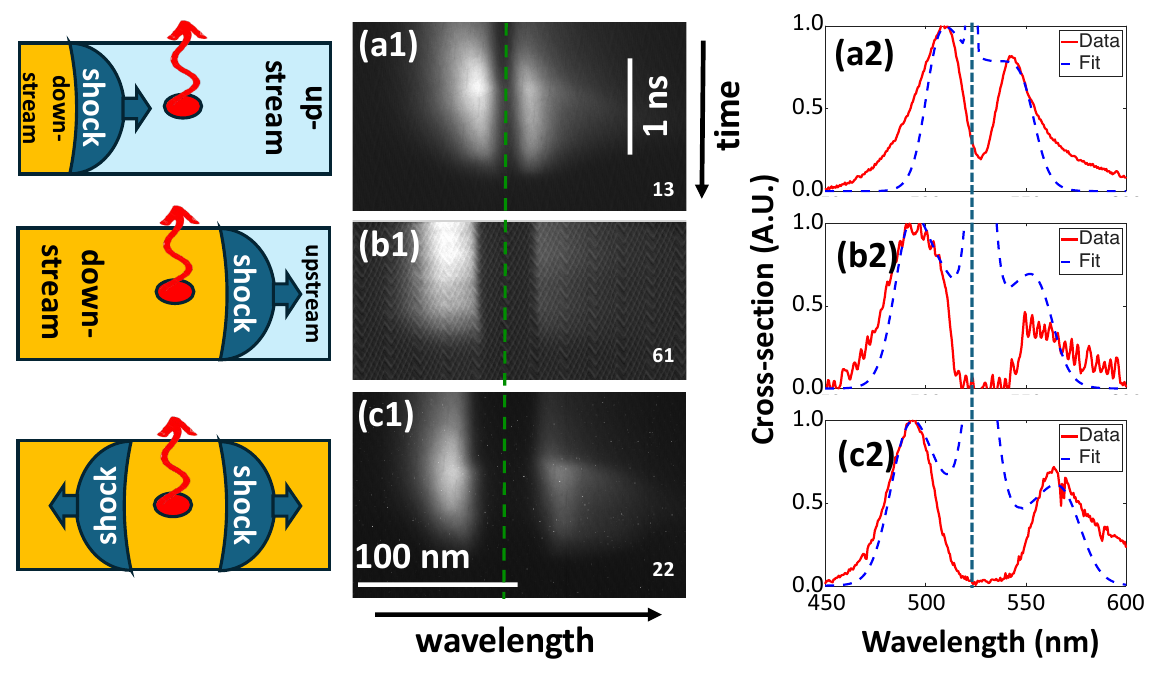}
    \caption{\textcolor{red}{Examples of TSe (see text for details). The patterns in (b1) are CCD readout noise. The signal is cut temporally in (b1) because of temporal jitter in the streak camera that shifted the signal upward in time. The green vertical dashed line marks the wavelength of the TS probe laser. A mask (with a variable width, depending on the shots) is positioned in front of the recording streak camera to block the unshifted TS laser light, resulting in the dark central gully. The cartoons on the left describe the situation of each probing: (a1) before shock crossing of the TS probed volume. (b1) After crossing by a single shock, and (c1) after crossing by the two shocks.}
    \textcolor{red}{Lineouts of the spectra shown in (a1)-(c1) across the wavelength. }
   \textcolor{red}{ The fit parameters are: for (a2), $T_e = 100~\mathrm{eV}$, $T_i = 10~\mathrm{eV}$, $n_e = (2 \pm 0.1) \times 10^{18}~\mathrm{cm}^{-3}$; for (b2), $T_e = 200~\mathrm{eV}$, $T_i = 100~\mathrm{eV}$, $n_e = (6.2 \pm 0.2) \times 10^{18}~\mathrm{cm}^{-3}$; for (c2), $T_e = 270~\mathrm{eV}$, $T_i = 180~\mathrm{eV}$, $n_e = (9.8 \pm 0.3) \times 10^{18}~\mathrm{cm}^{-3}$.}}
    \label{fig:TSe}
\end{figure*}

\textcolor{red}{Fig.~\ref{fig:TSe}a1-c1 below shows examples of TSe raw data, and Fig.~\ref{fig:TSe}a2-c2 shows the corresponding quantitative lineouts, with the theoretical best fits. These fits correspond to the theoretical equation of the scattered spectrum for coherent TS in unmagnetized and non-collisional plasmas, taking into account the instrumental function.}
\textcolor{red}{Fig.~\ref{fig:TSe}a2 corresponds to the probing (at $t=0.5~\mathrm{ns}$) of the ambient plasma, i.e., before the crossing of the shock. It shows the properties of the plasma upstream of the shock. Fig.~\ref{fig:TSe}b2 corresponds to the probing of the downstream plasma, after crossing by a single shock (at $t=2~\mathrm{ns}$). We see the associated increase in density of the plasma by a factor of around 3 (compare Fig.~\ref{fig:TSe}a2 and Fig.~\ref{fig:TSe}b2). Fig.~\ref{fig:TSe}c2 corresponds to the probing of the plasma, around the time the two shocks interpenetrate (at $t=1~\mathrm{ns}$). We see that the density has been further increased (compare Fig.~\ref{fig:TSe}a2 and Fig.~\ref{fig:TSe}c2, with respect to the upstream plasma, by a factor of around 5.}

\begin{figure*}
    \centering
    \includegraphics[width=0.7\linewidth]{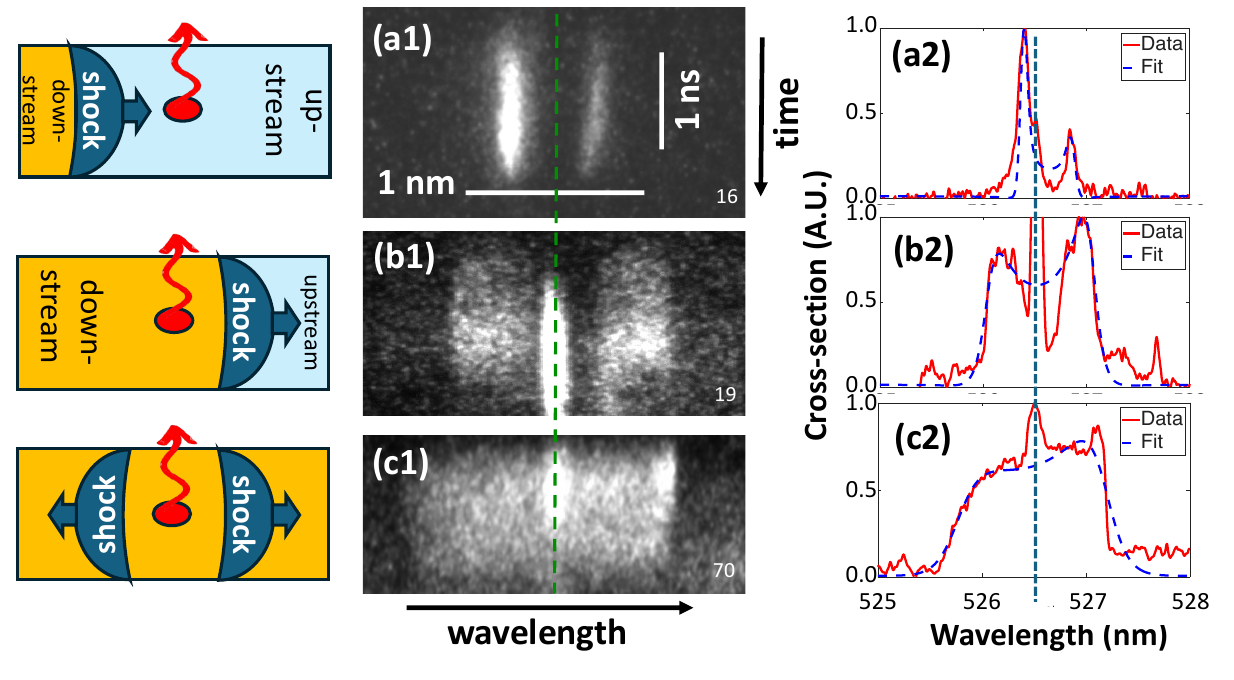}
    \caption{\textcolor{red}{Examples of TSi (see text for details). The green vertical dashed line marks the wavelength of the TS probe laser. The cartoons on the left describe the situation of each probing: (a1) before shock crossing of the TS probed volume. (b1) After crossing by a single shock, and (c1) after crossing by the two shocks. }
   \textcolor{red}{ Lineouts of the spectra shown in (a1)-(c1) across the wavelength. The green vertical dashed line marks the wavelength of the TS probe laser. The fits parameters are: for (a2), $T_e = 90 \pm 10$ eV, $T_i = 10 \pm 2$ eV, $n_e = 2 \times 10^{18}$ $\text{cm}^{-3}$, for (b2), $T_e = 202 \pm 10$ eV, $T_i = 80 \pm 20$ eV, $n_e = 6 \times 10^{18}$ $\text{cm}^{-3}$, for (c2), $T_e = 250 \pm 30$ eV, $T_i = 200 \pm 30$ eV, $n_e = 10^{19}$ $\text{cm}^{-3}$.}}
    \label{fig:TSi}
\end{figure*}

\textcolor{red}{Complementarily, Fig.~\ref{fig:TSi}a1-c1 below shows examples of TSi raw data, and Fig.~\ref{fig:TSi}a2-c2 shows the corresponding quantitative lineouts, with the theoretical best fits. 
As for TSe, these cases exemplify the different plasma conditions. Fig.~\ref{fig:TSi}a2 corresponds to the probing (at $t=0.75~\mathrm{ns}$) of the upstream ambient plasma, i.e., before the crossing of the shock. Fig.~\ref{fig:TSi}b2 corresponds to the probing (at $t=1.3~\mathrm{ns}$) of the downstream plasma, around the time of the crossing of the TS probed volume by a single shock. We see that both the electron ($T_e$) and ion ($T_i$) temperatures increase, compared to those of the upstream plasma, respectively by a factor of about 2 and 8. Fig.~\ref{fig:TSi}c2 shows that the temperature (here recorded at $t=3~\mathrm{ns}$) further increases when the two shocks interpenetrate, with a very notable increase of $T_i$, as can be grasped by the fact that the individual ion peaks become very broad and almost indistinguishable. }

\begin{figure*}
    \centering
    \includegraphics[width=0.7\textwidth]{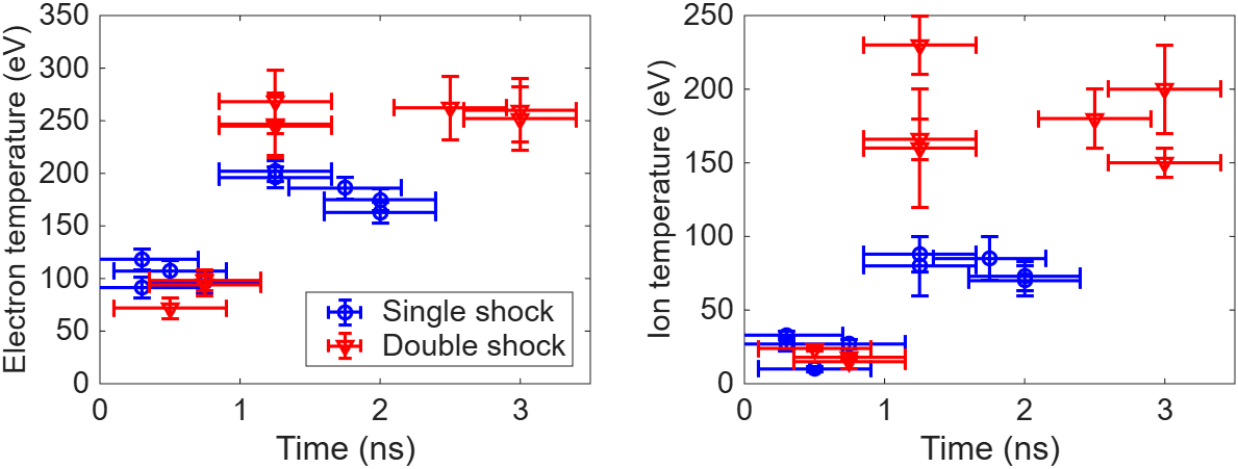}
    \caption{\textcolor{red}{Evolution of $T_e$ (left) and $T_i$ (right) as a function of time after the irradiation of the solid targets by the main lasers. The location of the volume probed by the TS diagnostic is such that the shocks arrive there at 1 ns, which is coincident with the observed temperature jump. The vertical error bar is related to the possible range of parameters in order to obtain a good fit of the data, as exemplified in Fig.~\ref{fig:TSe} and ~\ref{fig:TSi}, while the horizontal error bar is related to the duration of the TS probe beam.}}
    \label{fig:TS_summary}
\end{figure*}

\textcolor{red}{Fig.~\ref{fig:TS_summary} summarizes the temporal evolution of the electron and ion temperatures as inferred from TS probing of the plasma at various times.}

\textcolor{red}{Table~\ref{tab:TS_parameters} summarizes the plasma parameters for the three cases: upstream of the shocks, downstream of a single shock, and downstream of the two shocks after their interpenetration. 
The Rankine-Hugoniot jump conditions for a $\gamma = 5/3$ ideal fluid predict a compression ratio of 4 for a single strong shock. The slightly lower measured value of $\sim 3$ likely arises from the density corrugation visible at the shock front in the proton radiographs.
This corrugation causes local compression to depart from the ideal planar 1D picture, reducing the average compression across the corrugated shock surface relative to the ideal value.
The same departure from ideal 1D geometry applies to the colliding-shock case, where the cumulative jump of $\sim 5$ reflects realistic Mach numbers and corrugation effects, rather than the ideal product of two strong-shock compressions.}

\begin{table*}[ht]
\caption{\textcolor{red}{Average plasma parameters inferred from Thomson scattering measurements for the three different plasma configurations that are characterized: US, the ambient plasma upstream of the shock(s), $DS_S$, the ambient downstream plasma  of a single shock, and $DS_D$, the ambient downstream plasma,  after the two crossing shocks have passed through it.}}
\label{tab:TS_parameters}
\centering
\begin{tabular}{lccc}
\toprule
 & $\rm US$ & $\rm DS_S$ (Single) & $\rm DS_D$ (Double) \\
\midrule
$n_e$ (cm$^{-3}$) & $(2\pm 0.1)\times 10^{18}$ & $(6\pm 0.2)\times 10^{18}$ & $(1\pm 0.3)\times 10^{19}$ \\
$T_e$ (eV)        & $(100\pm 10)$ & $(200\pm 10)$ & $(270\pm 30)$ \\
$T_i$ (eV)        & $(10\pm 2)$  & $(100\pm 20)$ & $(200 \pm 30)$\\
\bottomrule
\end{tabular}
\end{table*}

\section{\textcolor{red}{Collisionality of the plasmas}}

\begin{figure}
    \centering
    \includegraphics[width=1.0\linewidth]{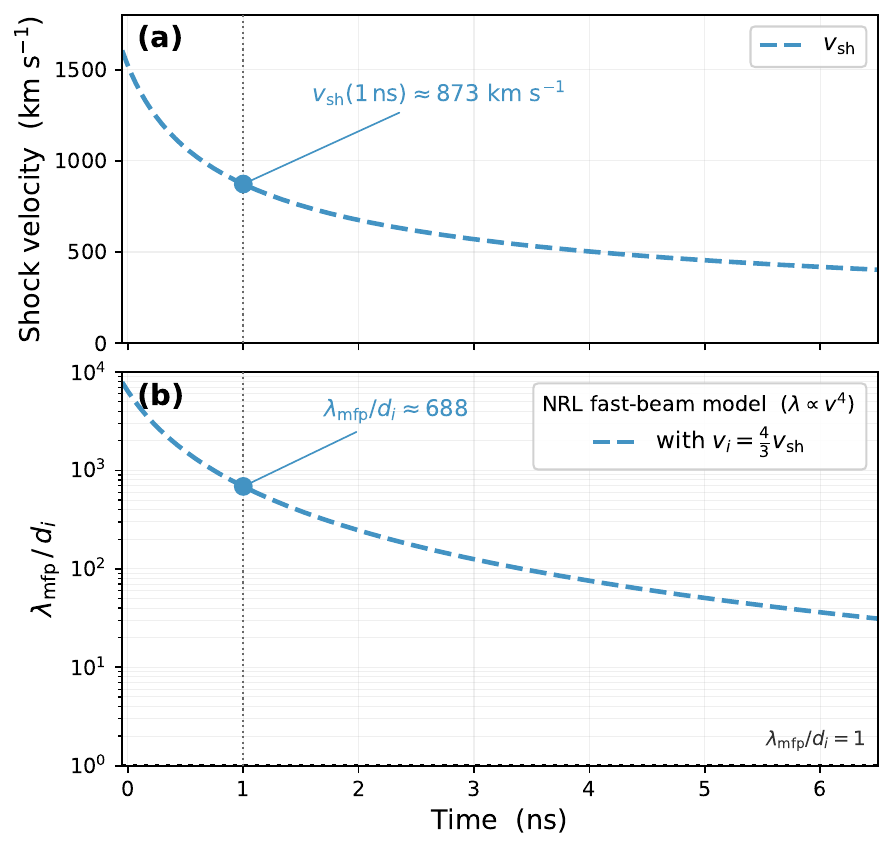}
    \caption{\textcolor{red}{\textbf{Time-resolved collisionality of the laboratory shock.}
    \textbf{(a)} Instantaneous shock velocity $v_{sh}$ as a function of
    time, derived from the shock expansion model fitted to the
    proton-radiography data (see Section~1).
    \textbf{(b)} Ion mean free path normalised to the ion inertial length,
    $\lambda_{\rm mfp}/d_i$, of a test proton incoming from the upstream and
    evaluated in the compressed downstream frame, where it moves at
    $v_i = (4/3)\,v_{sh}$ (since $n_d/n_u \approx 3$). The mean free path is
    computed from the NRL fast-test-particle transverse collision frequency
    ($\lambda_{\rm mfp} = v_i/\nu_\perp^{i/i'} \propto v_{sh}^{4}$) using the
    measured downstream conditions of the colliding-shock case
    ($n_d \approx 6\times10^{18}\,\mathrm{cm^{-3}}$, $\ln\Lambda \approx 10$);
    the field-ion temperature drops out in the fast-particle limit. The
    vertical dotted line marks $t = 1$~ns, the time at which the two
    counter-propagating shocks first interact. The curve in (b) satisfies
    $\lambda_{\rm mfp}/d_i \gg 1$ throughout the experimental window
    (1--6~ns), confirming that the plasma remains deeply collisionless
    during the experimentally relevant times.}}
    \label{fig:collision}
\end{figure}

\textcolor{red}{Regarding the collisionality, we work in the compressed downstream frame and evaluate the local ion--ion mean free path $\lambda_{\rm mfp}$ of a test proton incoming from the upstream. In this frame the test particle moves at an initial velocity $v_i = (1 + n_u/n_d)\,v_{sh}$, where $n_d/n_u$ is the downstream-to-upstream density ratio; since $n_d/n_u \approx 3$, we have $v_i = (4/3)\,v_{sh}$. The downstream plasma has an ion density $n_d \approx 6\times10^{18}\,\mathrm{cm^{-3}}$ and an ion temperature $T_d \approx 100\,\mathrm{eV}$, so that the ion thermal velocity $v_{th,d}$ is much smaller than the incoming velocity ($v_i \approx 12\,v_{th,d}$). We can therefore evaluate $\lambda_{\rm mfp} = v_i/\nu_\perp^{i/i'}$ using the NRL fast-test-particle transverse collision frequency \cite{NRL_Formulary}, $\nu_\perp^{i/i'} = 1.8\times10^{-7}\,n_d\, \ln\Lambda\, \varepsilon^{-3/2}$, with $\varepsilon = \tfrac{1}{2} m_i v_i^{2}$ the test-particle energy (in eV), $n_d$ in $\mathrm{cm^{-3}}$, and $\ln\Lambda \approx 10$ the Coulomb logarithm. Since $\nu_\perp^{i/i'} \propto v_i^{-3}$, one has $\lambda_{\rm mfp} \propto v_i^{4} \propto v_{sh}^{4}$; note that the field-ion temperature drops out in this fast-particle limit. For the reference velocity we take the inferred shock velocity at $t = 1\,\mathrm{ns}$, $v_{sh} \approx 873\,\mathrm{km\,s^{-1}}$, given by the expansion model fitted to the measured proton-radiography trajectory $R(t)$ within the experimental window (see Fig.~\ref{fig:collision}a and Section~1), with $t_{int} \approx 1\,\mathrm{ns}$ the time at which the two counter-propagating shocks first interact.}



\textcolor{red}{This yields $\lambda_{\rm mfp}/d_i \approx 690$ at $t = 1\,\mathrm{ns}$ (see Fig.~\ref{fig:collision}b and Table~1 of the main text), demonstrating the highly collisionless character of the shock; the same conclusion holds \emph{a fortiori} at earlier times, where the shock velocity is higher. As shown in Fig.~\ref{fig:collision}, the plasma remains deeply collisionless throughout the entire experimental observation window ($1$--$6\,\mathrm{ns}$). Although the shock will eventually transition to a collisional regime, this occurs well outside the experimental window. Moreover, this conclusion is unaffected for the energetic re-accelerated protons, which have higher velocities and propagate into lower-density regions; thus a later collisional transition of the bulk plasma would not influence their dynamics.}

\section{\textcolor{red}{Detailed proton spectra}}
\subsection{{\textcolor{red}{Proton spectra recorded in different conditions}}}
\begin{figure}
    \centering
    \includegraphics[width=0.45\textwidth]{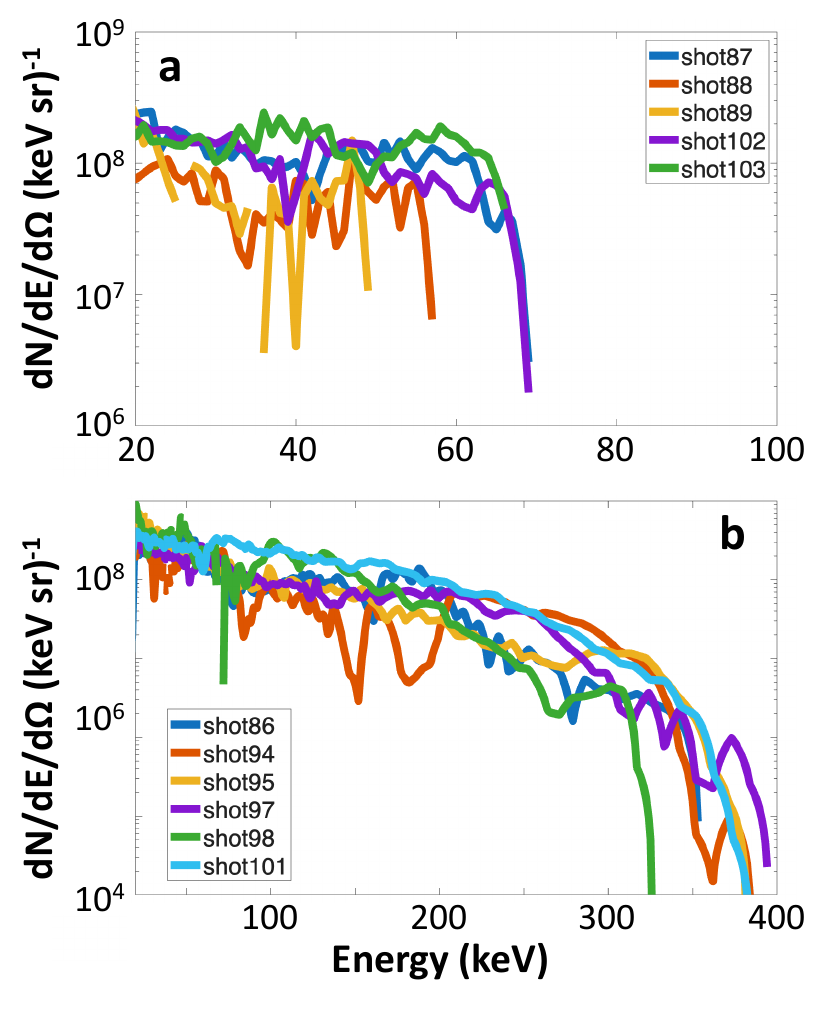}
    \caption{\textcolor{red}{a. Spectra of the protons exiting the plasma, as recorded by the spectrometer for 5 individual shots, showing the reproducibility of the acceleration recorded in the single shock case. b. Spectra of the protons exiting the plasma, as recorded by the spectrometer for 6 individual shots, showing the reproducibility of the acceleration recorded in the double shocks case. Note that the energy axis is much more extended here, compared to that of panel a}.}
    \label{fig:more_spectra}
\end{figure}

\textcolor{red}{The good reproducibility of the proton spectra collected along the axis of the external magnetic field can be seen on Fig.~\ref{fig:more_spectra}. Fig.~\ref{fig:more_spectra}a  shows the proton spectra recorded of several shots with only a single shock. We observe fluctuations, but that the maximum energy never goes beyond 70 keV, which is consistent with our previous results \cite{yao2021laboratory,yao2022detailed}, using the same platforms, namely at TITAN (LLNL) and LULI2000 (France), which shows also the reproducibility of the physical mechanism across experiments and lasers powering the plasma ablation. We emphasize also that we use the same material (Teflon) for the ablated plasma, the same gas jet, and most importantly the same external magnetic field coil, powering the large-scale magnetization of the plasma.}

\textcolor{red}{Fig.~\ref{fig:more_spectra}b shows the spectra of the 6 shots recorded in the double shock case. These were averaged to plot the averaged spectrum shown in the main paper, but we can see that the individual spectra are quite close from shot-to-shot to each other.
From the shot number, one can see also that we took care to perform single shock shots before and after with double shocks shots, to make sure that we could indeed confirm the energy boost observed when having two shocks - shots performed before and after those, with a single shock, did indeed not exhibit that energy boost.
}

\begin{figure}
    \centering
    \includegraphics[width=0.45\textwidth]{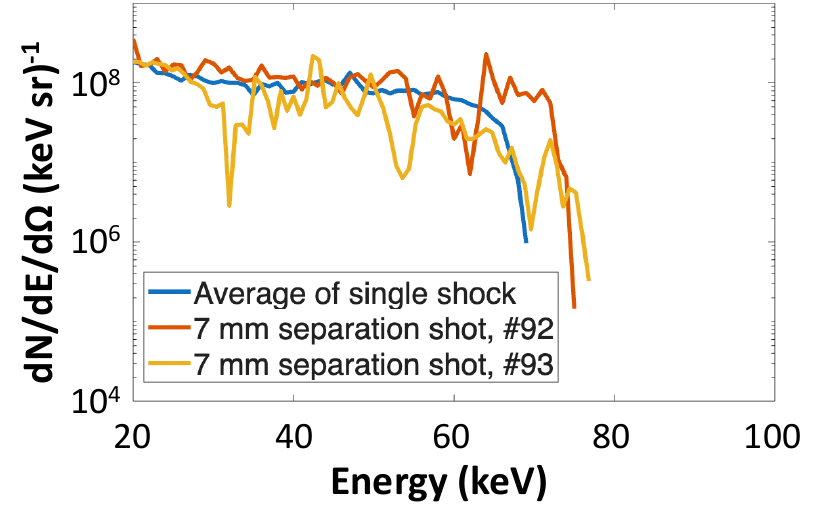}
    \caption{\textcolor{red}{Comparison of the proton spectra recorded when having a single shock (in blue - shown here is the average over 5 shots, the individual shots are shown in Fig.~\ref{fig:more_spectra}a) and of the two shots that were performed with two interacting shocks, with the target separation set at 7 mm (in red and orange).}
}
    \label{fig:more_spectra2}
\end{figure}

\textcolor{red}{We also performed two test shots with a target separation of 7 mm. The resulting proton spectra are shown in Fig.~\ref{fig:more_spectra2}, compared to the averaged single-shock spectrum. At this separation, only a modest enhancement in maximum proton energy is observed relative to the single-shock baseline. This is consistent with our former work \cite{fazzini2022particle,yao2023investigating}, where a large shock separation (9 mm) and sub-critical collision conditions also produced limited energy gains.}

\subsection{{\textcolor{red}{Discussion regarding the comparison between the measurements and the PIC simulations}}}

\textcolor{red}{
We first note that simulated and experimental proton spectra are not directly comparable in absolute terms: simulated spectra are snapshots of all particles within the simulation box at a given time, while experimental spectra are time-integrated and include only the subset of protons that escape along the applied magnetic-field ($z$) axis and reach the spectrometer placed at the edge of the coil. Diffusion, escape efficiency, and the finite detector acceptance therefore unavoidably reshape the measured distribution–particularly at low energy, where scattering losses are largest–so we do not claim a quantitative one-to-one correspondence in spectral shape.}

\textcolor{red}{
The comparison we draw is more targeted: both the 1D single-shock and the 1D/2D double-shock simulations reach cutoff energies consistent with the corresponding experimental measurements on a comparable timescale of $3$--$4$ ns. Thus, the relative energy boost between the single- and double-shock configurations observed experimentally is reproduced in the simulations. The 2D and 1D double-shock runs use different box and plasma sizes to reduce computational cost, so they are not expected to reach the same cutoff at the same time. The fact that the experimental energy boost is recovered across several simulation parameters and in both 1D and 2D supports our interpretation that the essential physical mechanism–reacceleration at the second shock crossing via the convective field $E_y = - B_z v_x$ – is captured by the simulations, regardless of the residual differences in spectral shape between 1D and 2D geometries (discussed separately in the manuscript). }

\section{\textcolor{red}{Discussion on 1D vs. 2D PIC simulations}}

\begin{figure}
    \centering
    \includegraphics[width=0.45\textwidth]{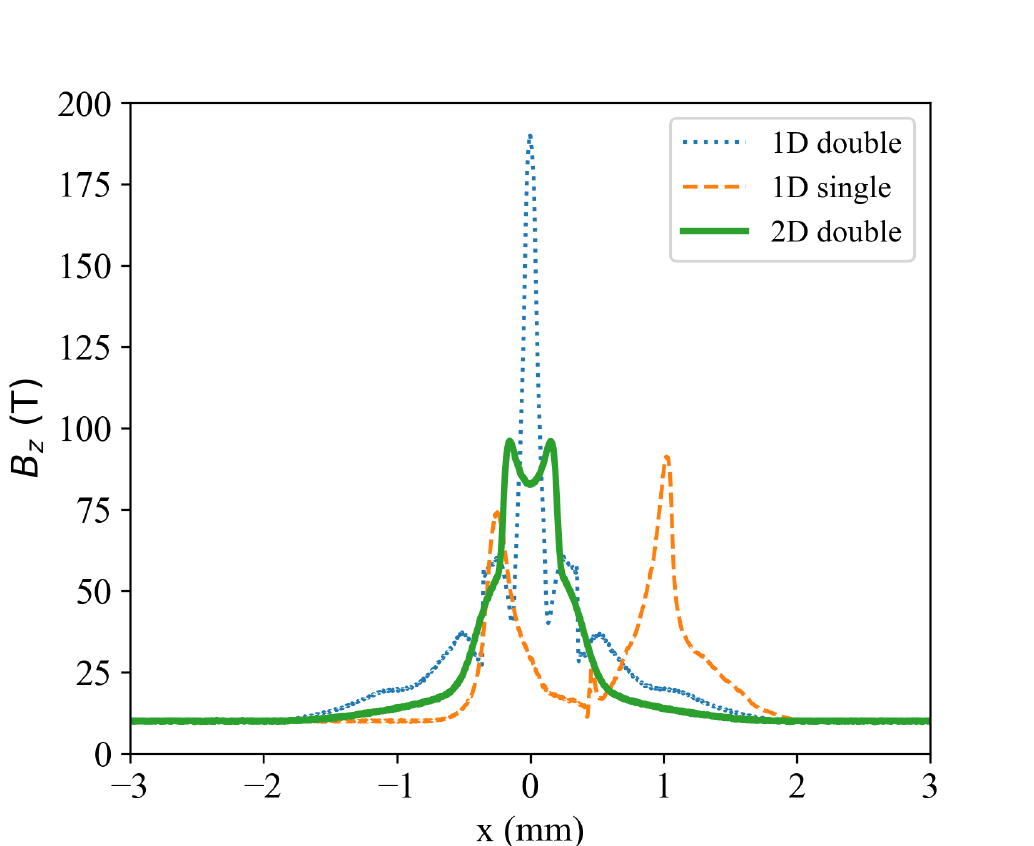}
    \caption{\textcolor{red}{Simulated profiles of the $B_z$ magnetic field  along the shock-normal direction ($x$), and at the time when the peaks of the $B_z$ compression reach their maximum , for the three simulation cases: 1D double shock (blue dotted), 1D single shock (orange dashed, showing the forward and reverse shocks), and 2D double shock (green solid).}}
    \label{fig:b_comp}
\end{figure}

\textcolor{red}{\textbf{Validity of the 1D approach:} 
The proton radiographs (Figs. 1 and 2) reveal the shock front as a curved interface with a radius of curvature $R_\mathrm{curv} \approx 2.7~\mathrm{mm}$ at $t = 0.95$ ns, and show that the local field at the merged shock reaches $B \approx 100~\mathrm{T}$ -- consistent with the 1D and the 2D PIC runs (Fig.~\ref{fig:b_comp}). The proton Larmor radius is $r_L = m_p v / (q B)$, giving $r_L \approx 0.26~\mathrm{mm}$ at the ambient velocity $v \approx 2500~\mathrm{km/s}$. Thus we have $r_L / R_\mathrm{curv} \approx 0.1$. Even at the observed cutoff energy ($\sim 400~\mathrm{keV}$),  the ratio remains relatively low ($\sim 0.3$), justifying the validity of the 1D approach. Curvature effects would dominate only for velocities above $v^* = \frac{q B\, R_\mathrm{curv}}{m_p} \approx 2.58 \times 10^{4}~\mathrm{km/s}$, corresponding to an energy $E^* = \tfrac{1}{2} m_p (v^*)^2 \approx 3.5~\mathrm{MeV}$, nearly an order of magnitude above the observed cutoff. Therefore, the shock front is seen as locally planar for the whole accelerated population, and multiple interactions with $E_y = B_z v_x$ are not geometrically forbidden.} 

\textcolor{red}{
On the other hand, the 2D PIC double-shock run recovers both the local field ($B \approx 100~\mathrm{T}$,Fig.~\ref{fig:b_comp}) and a proton cutoff energy comparable to the 1D simulation and the experiment (Fig. 3). Together, the ratio $r_L / R_\mathrm{curv} \approx 0.1$ at relevant proton energies and the agreement between the 1D/2D simulation and experimental spectra support the validity of the essentially 1D picture of the energy boost.}

\begin{figure}
    \centering
    \includegraphics[width=0.45\textwidth]{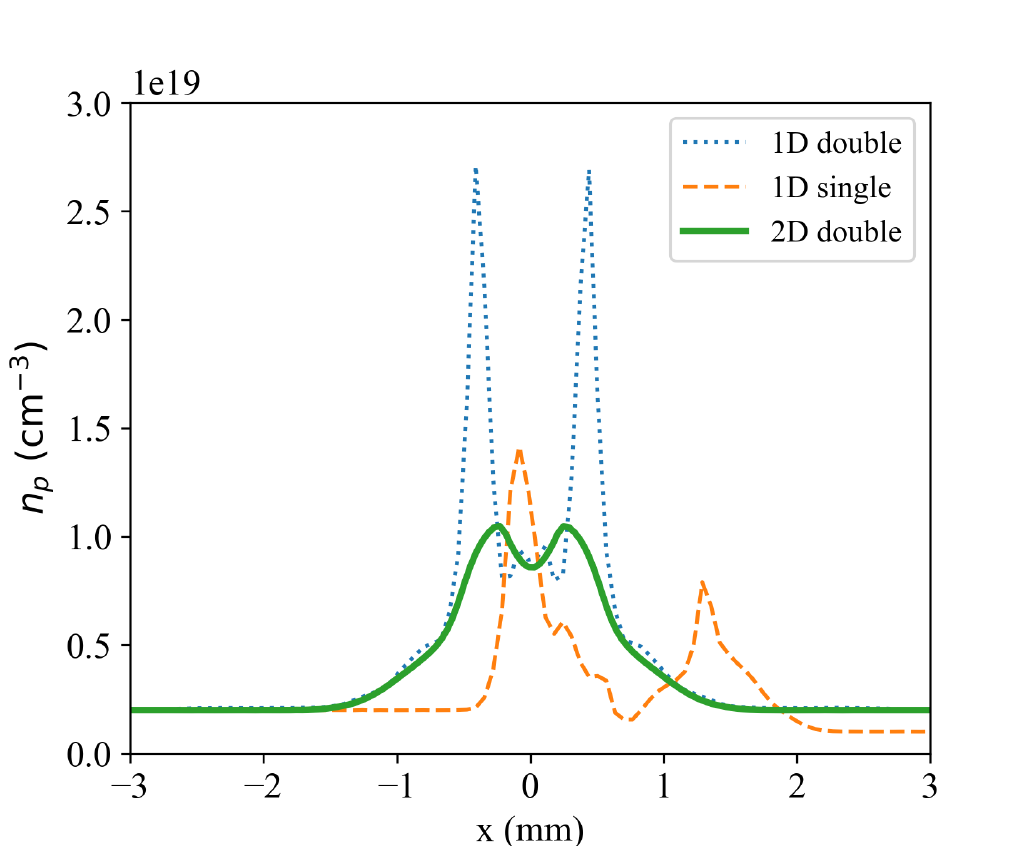}
    \caption{\textcolor{red}{Simulated proton number density profiles along the shock-normal direction ($x$), at the time when the peaks reach their maximum , for the three simulation cases: 1D PIC double shock (blue dotted), 1D PIC single shock (orange dashed, showing the forward and reverse shocks), and 2D PIC double shock (green solid). The upstream density is $1 \times 10^{18}~\mathrm{cm^{-3}}$. In the double-shock cases, the two shocks propagate symmetrically inward from both sides, leaving unshocked upstream plasma in the centre. The 1D PIC double-shock case reaches a peak compression of $\sim 27$, significantly higher than the 2D case ($\sim 10$), due to the absence of lateral expansion and reduced heating in 1D PIC geometry. The 2D PIC density profile is obtained by averaging over the transverse direction.}}
    \label{fig:dens_comp}
\end{figure}

\textcolor{red}{\textbf{Discussion of the density compression observed in 1D/2D simulation vs. the experimental one:} 
In the 1D double-shock case, the absence of lateral expansion and reduced heating (compared to higher-dimensional runs) result in a peak proton density of $n_p \approx 2.7 \times 10^{19}~\mathrm{cm}^{-3}$, corresponding to a compression ratio of $\sim 27$ relative to the upstream density of $n_{p,0} = 1 \times 10^{18}~\mathrm{cm}^{-3}$  (Fig.~\ref{fig:dens_comp}). 
The 2D double-shock simulation, which leads to stronger heating, predicts a lower maximum density of $\sim 1 \times 10^{19}~\mathrm{cm}^{-3}$ ($\sim 10\times$ compression), while a full 3D geometry would likely reduce it further. The Thomson scattering measurement gives an even lower density jump of $\sim 3$ (for single shock) and $\sim 5$ (for double shocks), see Table~\ref{tab:TS_parameters}, because the diagnostic is both spatially integrated over the probe volume and partially time-integrated: consequently, the high-density shock front -- spatially thin and temporally transient -- is diluted by the surrounding unshocked plasma.
}

\begin{figure}
    \centering
    \includegraphics[width=0.45\textwidth]{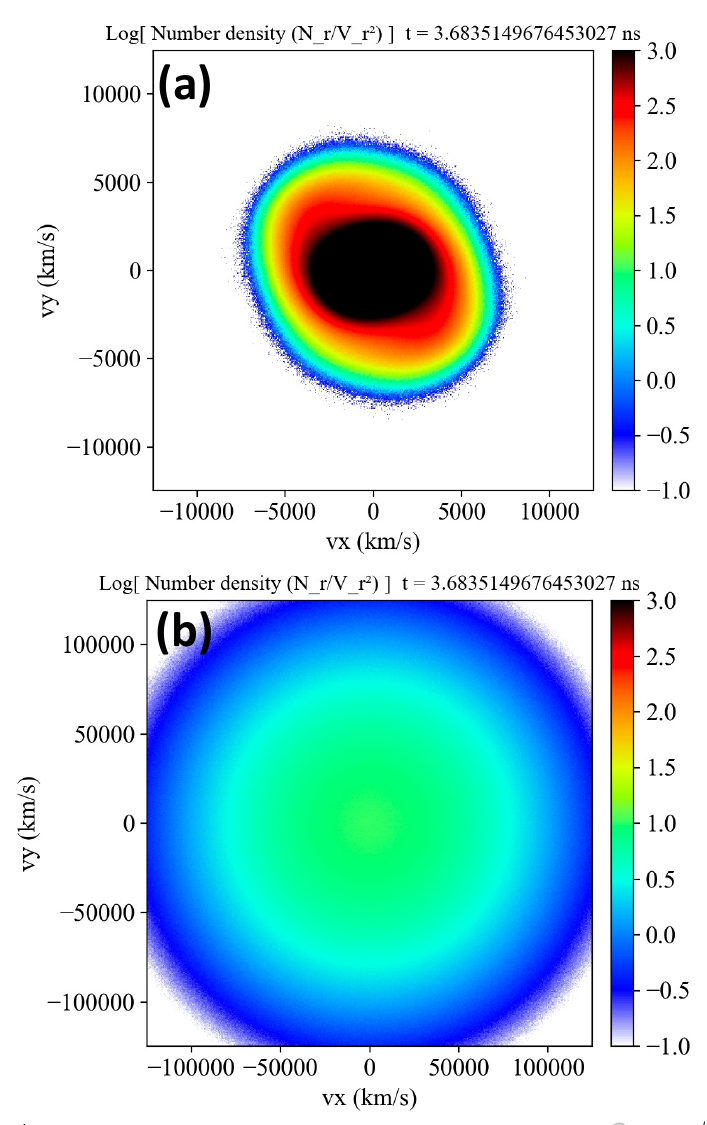}
    \caption{\textcolor{red}{Phase-space distribution ($v_x-v_y$) of (a) protons and (b) electrons from the 2D PIC simulations at 3.68 ns.}}
    \label{fig:px_py}
\end{figure}

\textcolor{red}{\textbf{Use of the 2D simulations for anisotropy evaluation:} 
Compared to 1D simulations, 2D simulations allow us to check whether there is a preferred acceleration direction within the plane perpendicular to the external B field. Figure~\ref{fig:px_py} shows the phase-space distribution $f(v_x, v_y)$ from the 2D PIC simulation at $t = 3.68$ ns. The proton distribution (panel a) exhibits no pronounced anisotropy, suggesting roughly isotropic in-plane emission.}

\newpage
\bibliographystyle{naturemag}
\bibliography{main}